\documentclass[11pt]{amsart}

\usepackage{amsmath}
\usepackage{amssymb}
\usepackage{amsthm}
\usepackage{amsfonts}
\usepackage{amstext}
\usepackage{amsopn}
\usepackage{amsxtra}
\usepackage{mathrsfs}

\newcommand{\E}{\mathrm{e}}
\newcommand{\I}{\mathrm{i}}
\newcommand{\Eq}{\mathcal{E}}
\newcommand{\D}{\mathrm {d}}
\newcommand{\Z}{\mathbb{Z}}

\newcommand{\T}{\mathbb{T}}
\newcommand{\Hi}{\mathcal{H}}
\newcommand{\B}{\mathcal{B}}

\newcommand{\F}{\mathcal{Z}}
\newcommand{\M}{\mathcal{M}}

\newcommand{\A}{\mathcal{A}}
\newcommand{\Or}{\mathcal{O}}

\newcommand{\epsi}{\varepsilon}
\newcommand{\ph}{\varphi}

\newcommand{\indexd}{\in\{ 1,\ldots,d \}}

\newcommand{\be}{\begin{equation}}
\newcommand{\ee}{\end{equation}}

\newcommand{\1}{\mathbf{1}}

\newcommand{\Hf}{\mathcal{H}_{\mathrm{f}}}      

\newcommand{\la}{\lambda}
\newcommand{\Base}{B}

\DeclareMathOperator{\im}{Im}
\DeclareMathOperator{\Tr}{Tr}    
\DeclareMathOperator{\tr}{tr}    
\DeclareMathOperator{\re}{Re}
\DeclareMathOperator{\ran}{Ran}

\DeclareMathOperator{\p}{\textbf{P}}

\newtheorem{lemma}{Lemma}[section]
\newtheorem{theorem}[lemma]{Theorem}
\newtheorem{proposition}[lemma]{Proposition}
\newtheorem{assumption}[lemma]{Assumption}

\theoremstyle{definition}

\theoremstyle{definition}
\newtheorem{remark}[lemma]{Remark}
\theoremstyle{definition}

{\catcode `\@=11 \global\let\AddToReset=\@addtoreset}
\AddToReset{equation}{section}

\newcommand{\N}{{\mathbb N }}
\newcommand{\R}{{\mathbb R}}
\newcommand{\C}{{\mathbb C}}
\newcommand{\e}{{\varepsilon }}
\newcommand{\ie}{{\sl i.e.\ }}
\newcommand{\cf}{{\sl cf.\ }}
\newcommand{\eg}{{\sl e.g.\ }}
\newcommand{\norm}[1]{ \left| \! \left| #1 \right| \! \right| }

\def\d{{\partial}}

\def\({\left(}
\def\){\right)}
\def\<{\left\langle}
\def\>{\right\rangle}
\def\O{\mathcal O}

\newcommand{\Id}[1]{{\rm 1\kern-2pt I_{#1}}}
\renewcommand{\hbar}{{\displaystyle\bar{\phantom{x}}\kern-6pt h}}

\numberwithin{equation}{section}

\begin{document}


\title[Piezoelectricity]{Geometric currents in piezoelectricity}

\author[G. Panati]{Gianluca Panati}
\address{Zentrum Mathematik, Technical University Munich, Boltzmannstrasse 3, D-85747 Garching, Germany
\newline
Dipartimento di Matematica, Universit\`{a} di Roma ``\,La Sapienza",
Piazzale Aldo Moro 2, I-00185, Roma, Italy}
\email{panati@ma.tum.de}

\author[C. Sparber]{Christof Sparber}
 \address{Wolfgang Pauli Institute Vienna \& Faculty of Mathematics, Vienna University, Nordbergstra\ss e 15, A-1090 Vienna, Austria}
\email{christof.sparber@univie.ac.at}

\author[S. Teufel]{Stefan Teufel}
 \address{Mathematics Institute,
 University T\"ubingen, Auf der Morgenstelle 10, D-72076 T\"ubingen, Germany}
 \email{stefan.teufel@uni-tuebingen.de}

\begin{abstract}
As a simple model for piezoelectricity we consider a gas of infinitely many
non-interacting electrons subject to a slowly time-dependent periodic potential. We
show that in the adiabatic limit the macroscopic current is determined by the
geometry of the Bloch bundle. As a consequence we obtain the King-Smith and
Vanderbilt formula up to errors smaller than any power of the adiabatic parameter.
\end{abstract}
\thanks{This work has been supported by the DFG Priority program ``Analysis, Modeling and
Simulation of Multiscale Problems''. C.S. has been supported by the APART research
grant funded by the Austrian Academy of Sciences.}

\maketitle


\maketitle

\section{Introduction}

In the year 1880 the brothers Jacques and Pierre Curie discovered that some
crystalline solids (like quartz, tourmaline, topaz, \ldots) exhibit a relevant
macroscopic property: if the sample is strained along a particular axis (called the
piezoelectric axis) a macroscopic polarization at the edges of the sample appears.

Even though first technological applications already appeared only a few years
later, a microscopic understanding of the phenomenon waited many decades after the
appearance of quantum mechanics. Up to the mid seventies, it was common lore that
the macroscopic (relative) polarization $\Delta \p = \p_{\rm fin} - \p_{\rm in}$
(\ie the polarization in the final state with respect to the initial state of the
sample) was due to the fact that, by deforming the crystal, the fundamental unit
cell acquires a \emph{non-vanishing electric dipole moment} with respect to the
unperturbed state. As pointed out by Martin in $1974$ \cite{Ma}, the previous
approach was intrinsically incorrect, since the total polarization should take into
account not only the sum of the dipole moments of the unit cells, but also the
\emph{transfer of charge} between unit cells. While in the ionic contribution
$\Delta \p _{\rm ion}$ the transfer of charge is negligible, it cannot be neglected
as far as the electronic contribution $\Delta \p _{\rm el}$ is
concerned.\footnote{\quad Thereby one clearly assumes that an approximate splitting
$\Delta \p = \Delta \p _{\rm ion} + \,\, \Delta \p _{\rm el}$ is justified.} 
It has thus been suggested by Resta \cite{Re92} to shift the attention from the
charge distribution (\ie the electric dipole moment) to the current, \cf the
review papers \cite{Re94, Resta_book2} and references given therein. In other words
one considers
$$
\Delta \p_{\rm el}  = \int_{T_{\rm in}}^{T_{\rm fin}} \!\!\!\D t \,\, \dot \p(t),
$$
where $\dot \p(t)$, called the \emph{piezoelectric current}, is the real quantity of
interest, see equation \eqref{Pcurr} below for the precise definition. Within this
framework, Resta used linear response theory in order to conveniently re-express
$\Delta\!\p_{\rm el}$ in terms of the Bloch functions \cite{Re92,Re94}.

Elaborating on Resta's result, King-Smith and Vanderbilt \cite{KSV} were able to relate the relative
polarization to the Berry connection, through the formula
\begin{equation}  \label{KSVa formula0}
\Delta \p_{\rm el}  =  \frac{1}{(2\pi)^d} \sum_{m=0}^{M} \int_{\T^*} \D k \,\, \big( \A_m(k,T) - \A_m(k,0) \big),
\end{equation}
where the sum runs over all the occupied Bloch bands, $d$ is the
space dimension, $\T^*$ denotes the first Brillouin zone, and
$\A_m(k,t)$ is the Berry connection for the $m$th Bloch band at time
$t\in \R$. Thereby the deformation is supposed to take place during
the time-interval $I=[0,T]$. The advantage of formula \eqref{KSVa
formula0} is twofold: it depends only on the occupied bands, and it
relates the macroscopic polarization to a geometric quantity, which,
as discussed later, does not depend on the particular gauge, \ie the
choice of the phase of the Bloch functions.

In this paper we provide a rigorous formula for $\Delta \p_{\rm el}$, which is more general than \eqref{KSVa formula0},
by exploiting the fact that the deformation of the
crystal is an adiabatic phenomenon, \ie it is extremely slow when measured on the
atomic time-scale. Moreover, we provide an alternative derivation of \eqref{KSVa
formula0} relying on the semiclassical dynamics of a state which is essentially
concentrated on a single isolated Bloch band.


\subsection{Description of the model}

In the following we shall focus only on the current induced by the electrons, which moreover are assumed
to be non-interacting. Thus $\Delta \! \p_{\rm el} \equiv \Delta \! \p$ to simplify the notation. Further we
shall restrict ourselves to the \emph{zero temperature regime}, thus taking into account only electrons
with an energy below the \emph{Fermi level} $E_*$.

The physical strain on the lattice will be modeled by a simple \emph{time-dependent
Hamiltonian} $H(t)$, to be specified below.  Here $t$ is  interpreted as the
\emph{macroscopic} time-scale which relates to the \emph{microscopic} time $s$ via
$t = \e s$, with $0<\e \ll 1$. In other words, a dimensionless small parameter
$\e\ll 1$ is introduced, describing the effects of the mechanical strain as slow
variations on the microscopic time scales. Thus we consider the asymptotic behavior
as $\e \to 0$ of  the following Schr\"odinger evolution system
\begin{equation}
\label{model1} \left \{
\begin{aligned}
\I\, \e \frac{{\rm d}}{{\rm d}t} \, U^\e(t,0) = & \, H (t) \, U^\e(t,0) ,\\[1mm]
U^\e(0,0)   = &  \, {\bf 1}_{\mathcal H},
\end{aligned}
\right.
\end{equation}
which consequently describes the dynamics of electrons on the macroscopic time scale $t=\e s$. For sake of
a simpler notation, we shall from now on write $U^\e(t)$ for $U^\e(t,0)$.

\medskip

Within this setting, the macroscopic polarization $\Delta \!
\p^{\e}$ is then defined as follows. The current operator (with
respect to \emph{macroscopic} time) is
\begin{equation}\label{curr}
J^\epsi :=      \frac{ \I }{\e} \,  [ H(t), x].
\end{equation}
In particular, in the case $H(t) = - \frac{1}{2}\Delta + V_{\Gamma}(t)$ considered
below, one has $ J^\epsi = -\frac{\I}{\epsi}\nabla_x$. Here, and in the following
all physical constants are set equal to 1 for convenience. We also assume that the
deformation of the solid takes place in a fixed {\em macroscopic} time interval
$I=[0,T]$, \ie that supp $\dot H(t)\subseteq I$. The state of the system at time $t$
is given by $\rho^\epsi(t) := U^\epsi(t)\, \rho(0)\,U^\epsi(t)^*$, where
$\rho(0):={\bf 1}_{(-\infty,E_*]}\( H(0) \)$ denotes the spectral projection of
$H(0)$ below a certain energy $E_*$, the Fermi energy. The macroscopic piezoelectric
current is thus defined as
\begin{equation} \label{Pcurr}
\dot {\p^{\e}}(t): = \mathcal T (\rho^\epsi(t)\, J^{\e}).
\end{equation} where $\mathcal T(A)$ denotes the so-called \emph{trace
per unit volume} of an operator A, \ie \begin{equation} \label{TraceperVolume}
\mathcal T (A) := \lim_{n\to\infty}\frac{1}{|\Lambda_n|} \re \Tr ({\bf
1}_{\Lambda_n} A)\,, \end{equation} and ${\bf 1}_{\Lambda_n}$ is the characteristic
function of a $d$-dimensional box with finite volume $|\Lambda_n|$, such that
$\Lambda_n \nearrow \R^d$. Clearly, the notion of trace per volume is needed since
$\rho^\epsi(t)$  is not trace class. In summary we get that the macroscopic
polarization is given by
\begin{equation}\label{defp}
\Delta \p^\e =  \int_{0}^{T} \!\!\!\D t \,\,\mathcal T (\rho^\epsi(t)\,J^\epsi)\, ,
\end{equation}
which will be the main object of our investigations.

The previous definitions correspond to the following physical picture. We are
considering a large system which, at each fixed macroscopic time, is in
thermodynamic equilibrium in the state $\rho(t)$. If $\mu(x)$ is a box, centered at
$x$, whose size is comparable with the lattice spacing, the microscopic current $
|\mu(x)|^{-1} \re\Tr \( \1_{\mu(x)} \, \rho(t) \, J^{\epsi}\) $ depends sensitively
on the microscopic position $x$. An average over a larger mesoscopic region
$\Lambda_{\rm meso}$ is needed in order to get rid of the microscopic fluctuations.
The use of the thermodynamic limit appearing in (\ref{Pcurr}) and (\ref{defp})
guaranties that $\dot {\p^{\e}}(t)$ and $\Delta \p^\e$ are indeed bulk properties of
the system, \ie independent of the actual size and shape of the test volume
$\Lambda_{\rm meso}$. For a real sample, the charge accumulated during the
deformation of the sample at a face $\Sigma$ is expected to be approximately
$\int_{\Sigma} \Delta \p^\e \cdot \, n_{\Sigma}$, where $n_{\Sigma}$ is the normal
vector to $\Sigma$.

\begin{remark}
While we shall discuss the trace per volume in a bit more detail later, let us remark here why \eqref{TraceperVolume}
is the correct definition, at least in the case of the current operator.
The current density associated with a Schr\"odinger wave function $\psi(x)$ is
$$
j^\epsi(x) := \frac{1}{\epsi}\im \overline{\psi}(x)\,\nabla\psi(x)= \re  \overline{\psi}(x)\,(J^\epsi\psi)(x)
$$
and thus the current in a region $\Lambda\subset\R^d$ is
\[
J^\epsi(\Lambda) =\re \int_\Lambda \D x \, \overline{\psi}(x)\,(J^\epsi\psi)(x)= \re\, \langle \psi, {\bf 1}_\Lambda J^\epsi\psi\rangle\,,
\]
which generalizes to $\re \Tr ({\bf 1}_\Lambda \rho J^\epsi)$ for general mixed
states $\rho$ and to \eqref{TraceperVolume} in the thermodynamic limit. One arrives
at the same formula by symmetrization of the localized current operator, \ie
$$
\re \Tr ({\bf 1}_\Lambda \rho J^\epsi) = \frac{1}{2}\Tr (\rho \,(J^\epsi {\bf
1}_\Lambda +  {\bf 1}_\Lambda J^\epsi)).
$$
Note that one can find other definitions of the current in a volume
within the literature, \eg $\Tr( {\bf 1}_\Lambda \rho J^\epsi)$, \ie
without taking the real part, or $\Tr (\rho  {\bf 1}_\Lambda J^\epsi
{\bf 1}_\Lambda)$, \ie by localizing the current operator through $
{\bf 1}_\Lambda J^\epsi {\bf 1}_\Lambda$. All these definitions
yield (presumably) the same thermodynamic limit and thus the same
macroscopic current.
\end{remark}

To describe the effects of strain upon the solid we consider the standard model
in the study of polarization effects, see \eg \cite{KSV} and \cite{Re92}, namely
the following time-dependent Hamiltonian on $\mathcal H=L^2(\R^d)$
\begin{equation}
\label{ham} H(t):=-\frac{1}{2}\Delta  + V_{\Gamma}(x,t).
\end{equation}
Since we aim to describe a crystalline structure, the potential $V_{\Gamma}\left(x,t \right)$ in
\eqref{ham} is assumed to be \emph{periodic}, for all $t \in I$, w.r.t. to some \emph{regular lattice}\footnote{%
\quad  We say that a set $\Gamma \subset \R^d$ is a \emph{regular lattice} if
$\Gamma$ is a maximal subgroup of the group $(\R^d,+)$. The requirement of a
group structure corresponds to the physical idea of composition of translations.}$^{,}$
\footnote{\quad Notice the distinction between the periodicity lattice $\Gamma$,
which is a lattice in the sense of the previous definition, and the ``atomic
lattice" representing the positions of the ionic cores, which generally is not.
\bigskip
} 
$\Gamma \simeq \mathbb Z^d$, \ie
\begin{equation}
V_{\Gamma}(x + \gamma,t) = V_{\Gamma}(x,t), \quad \forall \, x \in \R^d, \gamma \in \Gamma, \ t\in I.
\end{equation}
The centered \emph{fundamental domain} of $\Gamma$ is
$$
Y:=\left\{ x \in \R^d: \ x=\sum_{l=1}^d  \zeta_l \, \gamma_l, \ \zeta_l\in
\left[-\tfrac{1}{2} \, ,  \tfrac{1}{2}\right] \right\},
$$ where $(\gamma_1, \ldots, \gamma_d)$ are the generators of $\Gamma$.
The corresponding \emph{dual lattice} will be denoted by $\Gamma^*$ with centered
fundamental domain $Y^*$, usually called (first) \emph{Brillouin zone}. Also, we
shall use the notation $\T^*=\R^d/\Gamma^*$, \ie the $d$-dimensional torus induced
by $\Gamma^*$. In other words $\T^*$ denotes the Brillouin zone $Y^*$ equipped with
periodic boundary conditions.

\goodbreak

A model with a time-independent lattice might seem unrealistic at first glance.
However, many piezoelectric materials, \eg GaAs, exhibit  a crystallographic
structure in which the ``atomic lattice", representing the positions of the ionic
cores, consists of two sub-lattices corresponding to the two atomic species. Within
a good approximation, the deformation of the sub-lattices due to the external strain
can be neglected, and the only relevant effect of the strain is a relative
displacement of the two sub-lattices \cite{KSV}. This situation is mathematically
described by the model analyzed in this paper, \ie by a time-independent periodicity
lattice $\Gamma$ (fixed with respect to one of the two atomic sub-lattices) and a
time-dependent potential, which represents the change of the potential due to the
displacement of the other sub-lattice.


\subsection{Synopsis} Within the   framework described above we provide in Theorem~\ref{th1} a rigorous
justification and generalization of the King-Smith and Vanderbilt formula  \eqref{KSVa formula0}. We show, in particular, that if $V_\Gamma$ is
$C^{N+1}$, as a map from $I=[0,T]$  to $\B(H^2(\R^d),L^2(\R^d))$, then
\begin{equation}\label{Intro KSVa1}
\Delta \p^\e =  -\frac{1}{(2\pi)^d} \int_{0}^{T} \!\!\!\D t \int_{\T^*} \D k \,\, \Theta (k,t) + \Or(\epsi^{N})\, ,
\end{equation}
where
\begin{equation}\label{theta}
\Theta(k,t):=-\I \,\tr \left( P(k,t)\,[\partial_t P(k,t),\,\nabla_k P(k,t)\,]\,\right)\,,
\end{equation}
and $P(k,t)$ is the Bloch-Floquet fiber decomposition\footnote{ \, A brief summary
of Bloch-Floquet theory is provided in Section~\ref{Sec BlochFloquet}.} of the
spectral projector $P(t)={\bf 1}_{(-\infty,E(t)]}\(H(t)\)$ (the definition of $E(t)$
is given in Assumption~\ref{assgap}). Here and in the following the symbol $\tr$
denotes the trace in the fiber Hilbert space, namely $L^2(Y)$, see Section \ref{Sec
BlochFloquet}. Whenever all Bloch bands within $\ran P(k,t)$ are isolated, formula
\eqref{Intro KSVa1} implies \eqref{KSVa formula0}, up to an an error of order
$\Or(\e^{N})$. Note however that \eqref{Intro KSVa1} is more general, since it can
be applied also to situations where band crossings occur. One key ingredient in the
rigorous derivation of \eqref{Intro KSVa1} is the super-adiabatic expansion of the
fiber decomposition of the time-evolved Fermi projector $\rho^\epsi(k,t)$. For fixed
$k\in\T^*$ we use the standard super-adiabatic expansion developed by Nenciu
\cite{Ne}. However, since we need to differentiate with respect to $k$, as suggested
by formula \eqref{theta}, the expansion needs to be done uniformly on spaces of
equivariant functions.

In addition to \eqref{Intro KSVa1}, we also provide a dynamical understanding of the
same formula based on first order corrections to the semiclassical model of solids.
This comes at the price of restricting ourselves to the situation without band
crossings. In Theorem~\ref{th2} below we show that the semiclassical equations of
motion for an electron in the $m$th Bloch band, including $\Or(\e)$ corrections, are
\begin{equation}
\label{Intro sceq} \left \{
\begin{aligned}
& \dot q = \nabla_k E_m(k,t) -  \e\, \Theta_m(k,t), \\
& \dot k = 0,
\end{aligned}
\right.
\end{equation}
with $q$ being the macroscopic position and $k$ the crystal-momentum of the
electron. Here  $\Theta_m$ admits the representation
\begin{equation}\label{Piezocurvature2}
\Theta_m(k,t)= -\partial_t \A_m(k,t) - \nabla_k \phi_m(k,t),
\end{equation}
where one  introduces the geometric vector potential (Berry connection)
\begin{equation}\label{Berry connection}
\A_m(k,t)= \I \< \ph_m(k,t) , \nabla_k \ph_m  (k,t) \>_{L^2(Y)},
\end{equation}
and the geometric scalar potential
\begin{equation}\label{geometric potential}
\phi_m(k,t) = - \I \< \ph_m(k,t) , \d_t \ph_m (k,t) \>_{L^2(Y)},
\end{equation}
with $\ph_m$ being the $m$th Bloch eigenfunction. As suggested by the previous formulae,
the vector field $\Theta_m$ exhibit an interesting analogy with the electric field.
Moreover both $\Theta$ and $\Theta_m$ correspond to the curvature of a connection on
a bundle over $\T^* \times \R$, \cf Section~\ref{sgeo} for a
broader discussion on this.
It is then natural to baptize $\Theta$ the \emph{piezoelectric curvature}.

As shown in Section~\ref{Sec mainresults}, the King-Smith
and Vanderbilt formula \eqref{KSVa formula0} follows from the corrected semiclassical equations of motion \eqref{Intro sceq} by a
straightforward classical statistical mechanics argument. Indeed, it a standard textbook
argument which shows that the semiclassical model without the $\Or(\e)$
corrections implies that filled band do not contribute to the current at all.

\begin{remark} There is a  related result by Elgart and Schlein \cite{ElSch} who derive the
adiabatic charge transport for a class of Landau type Hamiltonians. They also rely
on Nenciu's super-adiabatic approximation to the time evolved Fermi projector. Here
we only remark that there are important differences between our result and
\cite{ElSch}. Details are given in the remarks after the statement of
Theorem~\ref{th1}.
\end{remark}
Our methods also apply to the case of a periodic deformation of the crystal, \ie
$H(t+T) = H(t)$ for every $t \in \R$. In such case, formula (\ref{Intro KSVa1})
implies that $\Delta \! \p^\e$ is, up to errors of order $\Or(\e^N)$, an integer
multiple of a fundamental quantity, in agreement with a previous observation by
Thouless \cite{Th}. Further analysis is required to show that $\Delta \! \p^\e$ is
actually nonzero in a specific model, as done in \cite{ABL} for the case of
Harper-like models. In general, in order to obtain a nonzero polarization one has to
choose a map $t \mapsto H(t)$ that, in a suitable space $\M$ of hamiltonian
operators, describes a loop around a manifold $\M_{\rm cr} \subset \M$ consisting of
hamiltonian operators for which the gap assumption (Assumption \ref{assgap}) is
violated. This fact is crucially used in \cite{ABL}, while an analogous situation
has been investigated in the context of molecular physics \cite{FaZi}.

\medskip

The paper is now organized as follows. The precise assumptions and the main mathematical results are stated in
Section~\ref{Sec mainresults}. In Section~\ref{sprel} we collect some preliminary results used in the
following. In Section~\ref{super} we present the so-called super-adiabatic theorem, which comprises the
main mathematical step towards our final results, to be proved in Section~\ref{sproof}. In
Section~\ref{sgeo} we discuss in more detail the geometrical interpretation of our results.

\bigskip

\textbf{Acknowledgements: }
We are grateful to M.\ Lein, U.\ Mauthner, H.\ Spohn and R.\ Tumulka for useful  comments and remarks.

\section{Main results} \label{Sec mainresults}

The basic assumption on the potential $V_\Gamma$ will be as follows.
\begin{assumption}\label{asspot}
For all $t\in \R$ the potential $ V_\Gamma(t)$ is $H_0$-bounded with relative bound
smaller than $1$. We assume that
$$ V_\Gamma\in
C^{N+1}(\R,\B(H^2(\R^d),L^2(\R^d)))$$ for some $N \in \N$, and that $\dot
V_{\Gamma}(t)$ is compactly supported in a bounded interval $I=[0,T]$ and $H(t)$-form bounded for all $t\in I$.
\end{assumption}
 From this assumption it follows in particular that $H(t)$ is self-adjoint on the Sobolev space $H^2(\R^d)$, for all $t\in
I$. Moreover this implies the existence of a unique unitary propagator $U^\e(t)$ obeying \eqref{model1}.

 From now on we  impose the following condition on the spectrum of $H(t)$:
\begin{assumption}
\label{assgap}
There exists a continuous function $E(t)$, such that $E(0)=E_*$, which satisfies
$$
\mbox{\emph{dist}}(E(t), \sigma(H(t)))>0,\quad \mbox{for all $t\in I$}.
$$
\end{assumption}
It is not assumed, however, that there is an energy $E_*$, independent of time, which lies in a
spectral gap for all $t\in I$, \ie the gap might move up and down in energy.

\begin{theorem} \label{th1}
Let Assumption \ref{asspot} hold and let $P(k,t)$ be the Bloch-Floquet
representation of the spectral projector $P(t)={\bf 1}_{(-\infty,E(t)]}\(H(t)\)$, where $E(t)$ is as in
Assumption \ref{assgap}. Then there exists an orthogonal projector $P^\e_N(k,t)$ with
$$
\|P(k,t)-P^\e_N(k,t)\| = \Or(\epsi)\,,
$$
such that the macroscopic current can be expressed as
\be\label{mres}
\begin{split}
\mathcal T( \rho^\epsi(t) \,J^\e) = & \, -\frac{1}{(2\pi)^d}\int_{\T^*} \D k\, \Theta_N^\e (k,t) +\Or(\epsi^{N})\\
= & \, -\frac{1}{(2\pi)^d}\int_{\T^*} \D k\, \Theta (k,t) +\Or(\epsi)\,,
\end{split}
\ee
where $\Theta(k,t)$ is given by \eqref{theta} and $\Theta^\e_N (k,t)$ is
\begin{equation}\label{thetaN}
\Theta^\e_N (k,t):=-\I \,\tr \left( P^\e_N(k,t)\,[\partial_t P^\e_N(k,t),\,\nabla_k
P^\e_N(k,t)\,]\,\right).
\end{equation}

The total transported charge is then
\begin{equation}
\label{poleps}
\begin{split}
\Delta \mbox{\emph{$\p^\e$}}  := & \  \int_{0}^{T} \!\!\!\D t \,\,\mathcal T (\rho^\epsi(t)\,J^\e)  = -\frac{1}{(2\pi)^d}\int_{0}^{T} \!\!\!\D t \int_{\T^*} \D k\, \Theta (k,t) +\Or(\epsi^{N})\, .
\end{split}
\ee
\end{theorem}

\goodbreak

{\bf Remarks}
\begin{enumerate}
\item Note that the error estimate in \eqref{poleps} is  better than what one would
naively guess from \eqref{mres}. As mentioned before, the King-Smith and Vanderbilt
formula  \eqref{KSVa formula0} was originally derived using linear response theory.
Our result \eqref{poleps} confirms the rule that there are no power law corrections
to Kubo's formula, see also \cite{KlSe}. \item  Note that $\Theta(k,t)$ is well
defined \emph{independently} of whether there are energy level crossings within the
occupied Bloch bands, \ie within $\ran P(k,t)$, or not, and independently of whether
the complex vector bundle defined by $\ran P(k,t)$, for fixed $t\in I$, is trivial
or not. \item  From the physical point of view, Assumption~\ref{assgap} corresponds
to the requirement that the solid remains an insulator during the adiabatic
deformation of the crystal, a condition verified in the experiments. Mathematically
one could relax Assumption~\ref{assgap} to hold only locally on the first Brillouin
zone. More precisely, Theorem~\ref{th1} is still valid by the same proof, if there
exist a continuous function $E(k,t)$, such that $E(k,0)=E_*(k)$, which satisfies
$$
\mbox{dist}(E(k,t), \sigma(H(k,t)))>0,\quad \mbox{for all $t\in I$ and $k\in \T^*$}.
$$
\item As remarked before, our result looks and to some extent is similar to the derivation of Kubo's formula for Landau type
Hamiltonians by Elgart and Schlein \cite{ElSch}. Therefore we would
like to explicitly point out some crucial differences. In
\cite{ElSch} only the leading order expression for the current is
computed. While it can be seen from \eqref{poleps} that the leading
order expression for the total charge is valid up to errors of order
$\epsi^{N+1}$, this is not true for the current itself, which is as
well an observable quantity. Another difference is that we are
looking at a bulk property, the macroscopic current, while in
\cite{ElSch} the authors consider the current induced in a fixed
finite region. On the other hand, since we heavily use the
periodicity of the problem, we can't allow for small but
non-periodic perturbations of the Hamiltonian so easily, as is done
in \cite{ElSch}.
\end{enumerate}

If all Bloch bands within $\ran P(k,t)$ are isolated, then
$\Theta(k,t)$ can be decomposed as
\begin{equation}\label{decom}
\Theta(k,t) = \sum_{m=0}^ M  \Theta_m(k,t), \qquad M = \dim \( \ran P(k,t)\),
\end{equation}
where $\Theta_m$ admits the representation \eqref{Piezocurvature2}. In particular \eqref{decom}, together
with \eqref{poleps}, gives
\begin{equation}  \label{KSVa formula}
\Delta \p^\e  =   \frac{1}{(2\pi)^d} \sum_{m=0}^{M} \int_{\T^*} \D k \,\, \big( \A_m(k,T) - \A_m(k,0) \big) + \mathcal O(\e^{N+1}),
\end{equation}
where the contribution of $\phi_m(k,t)$ vanishes due to periodicity, thus yielding a rigorous
justification of the King-Smith and Vanderbilt formula  \eqref{KSVa formula0}.

In our second result we give an alternative derivation of the King-Smith and Vanderbilt formula based
on first order corrections to the the semiclassical model in solids.
To this end we restrict ourselves to the case of a simple isolated Bloch band $E_m(\cdot, t)$, with the corresponding
eigenprojector denoted as $P_m(\cdot, t)$.
For sake of a simple discussion, we consider pure state solutions to the Schr\"odinger equation,
\begin{equation}
\psi^\e(t,x)=U^\e(t)\,\psi_0(x),\quad \psi_0\in \mbox{$\ran P_m(0)$},
\end{equation}
where the unitary propagator $U^\e(t)$ solves \eqref{model1}.  However the result extends
without major difficulties to the case of a mixed state, provided it is initially concentrated on the $m$-th
Bloch band, \ie $\rho_0 = P_m(0) \rho_0 P_m(0)$.

Since we are interested in the macroscopic charge distribution only,
we study the corresponding \emph{macroscopic Wigner function}  defined as
\begin{equation} \label{wigner1}
w[\psi^\e(t)](q,k):=\frac{1}{(2\pi \e)^d}\int_{\R^d} \, \D \eta\,
\overline{\psi^\e}\left(t,\frac{q}{\e}+\frac{\eta}{2}\right) \psi^\e\left(t,\frac{q}{\e}-\frac{\eta}{2}\right) \E^{\I \eta \cdot k}.
\end{equation}
The variable $q:=\epsi x$ will be called  the macroscopic position and $k\in \R^d$.
The Wigner function is the quantum mechanical analogue of the phase space
distribution in classical statistical mechanics, even though $w[\psi^\e(t)]$ is not
positive in general (for more details on Wigner functions we refer to \cite{Fol}).
Since the natural phase space in our case is $\R^d\times \T^*$ rather than
$\R^d\times\R^d$,  we  fold the Wigner function onto  $\R^d\times \T^*$ and  use the
\emph{reduced Wigner function} \cite{TePa} given by
\begin{equation}
w_{\rm r}[\psi^\e(t)](q,k)= \sum_{\gamma^*\in \Gamma^*} w[\psi^\e(t)](q,k+\gamma^*).
\end{equation}
The main use of $w_{\rm r}[\psi^\e(t)]$ is that it allows to compute expectation
values of  \emph{Weyl quantized operators} $a^W(\e x,-\I\nabla_x)$, \cf \cite{Fol,
Te} with semiclassical symbols $a\in C^\infty_{\rm b}(\R^6)$, such that
$a(q,k+\gamma^*)=a(q,k)$ for all $\gamma^* \in \Gamma^*$, via the following formula
\begin{equation} \label{wigner2}
\<\psi^\e(t), a^W \, \psi^\e(t)\>_{L^2(\R^d)} = \iint_{\R^d\times \T^*} \D q \, \D k
\ a(q,k) \, w_{\rm r}[\psi^\e(t)](q,k)\, .
\end{equation}
Above, $C^\infty_{\rm b}$ denotes the space of smooth functions which are bounded
together with all their partial derivatives.

The following theorem states that the Wigner function of the solution to the
Schr\"odinger equation can be approximately obtained by transporting the Wigner
function of the initial datum along the flow lines of a classical flow on
$\R^d\times\T^d$, the so called semiclassical model, \cf \eqref{sceq}.
\begin{theorem} \label{th2}
Let the potential $V_\Gamma$ satisfy Assumption \ref{asspot} with $N=1$. Also let
$E_m(k,t)$ be an isolated, non-degenerated Bloch band for all $t \in I$ and denote
by $P_m(t)$ the corresponding eigenprojector. Then, for any semiclassical observable $ a^W$, corresponding to a symbol $a\in
C^\infty_{\rm b}(\R^{2d})$ such that $a(x,k+\gamma^*)=a(x,k)$ for all $\gamma^*\in
\Gamma^*$, there is a constant $C_a$ such that  for any $\psi_0\in \ran P_m(0)$
and it holds
\begin{equation*}
\left|\iint\limits_{\,\,\,\,\R^d\times \T^*} \D q \, \D k \ a(q,k)\big(w_{\rm r}[\psi^\e(t)]-w_{\rm
r}[\psi_0]\circ \Phi_m^\e(0,t)\big)(q,k)\right| \leq \e^2  C_a \,|t|(1+|t|),
\end{equation*}
where   $\Phi_m^\e(t,0):\R^6\rightarrow \R^6$ denotes the $\e$-corrected semiclassical
flow in the $m$th Bloch band, given by the solution flow of
\begin{equation}
\label{sceq}
\left \{
\begin{aligned}
& \dot q = \nabla_k E_m(k,t) -  \e\, \Theta_m(k,t),\\
& \dot k = 0.
\end{aligned}
\right.
\end{equation}
\end{theorem}

In (\ref{sceq}) one interprets $q(t)$ as the macroscopic position and $k(t)$ as the
crystal-momentum of the electron.


The semiclassical equations of motion \eqref{sceq} can now be taken
as a starting point for a classical statistical mechanics analysis
of transport properties, \cf \cite{XSN}. Although the equations of
motion \eqref{sceq} are non-autonomous, they still allow for a
stationary measure. The stationary measure for a filled band at zero
temperature and with density of one particle per unit cell is
equi-distribution on phase space $\R^d\times \T^*$ with density
$(|Y||Y^*|)^{-1}=(2\pi)^{-d}$. The macroscopic current at time $t\in
I$ contributed from such a filled band is then
\begin{eqnarray*}
j_m^\epsi(t) &=& \frac{1}{\epsi (2\pi)^d}\int_{\T^*} \D k \, \dot q(k,t) \\
& =&  \frac{1}{\epsi (2\pi)^d}\int_{\T^*} \D k\, (\nabla_k E_m(k,t) -  \e\, \Theta_m(k,t))\\
&=& -\frac{1}{ (2\pi)^d}\int_{\T^*} \D k\,   \Theta_m(k,t)\,.
\end{eqnarray*}
Integration over the relevant time interval and summation over all filled bands
yields again the correct formula \eqref{poleps} for the polarization. Hence one can
understand the piezoelectric current even quantitatively on the basis of the
semiclassical model if one takes into account first order corrections. The situation
is similar to the quantum Hall current, \cf \cite{NTW, PST}.

\section{Technical preliminaries}\label{sprel}

To obtain a more detailed description of the properties of $H(t)$, we shall extensively use the well known
\emph{Bloch-Floquet theory}, some basic facts of which will be recapitulated in the next subsection. More
precisely, we will use a variant of the Bloch-Floquet transform which is sometimes called the Zak tranform
\cite{Zak}. A comparison of the two definitions is given in \cite{PST}.

\subsection{The Bloch-Floquet representation and the trace per volume} \label{Sec BlochFloquet} One exploits the periodicity of the
problem in order to separate the dynamics at the microscopic scale from the long
range dynamics. Denoting by $\T^d\equiv \R^d/\Gamma$ the $d$-dimensional torus
(corresponding to the fundamental cell $Y$ equipped with periodic boundary
conditions), the Bloch-Floquet-Zak transform
$$
\F: L^2(\R^d_x)\cong L^2(\Gamma \times Y)\cong \ell ^2(\Gamma)\otimes L^2(Y)\rightarrow L^2(Y_k^*, |Y^*|^{-1}\D k)\otimes L^2(\T^d_y)
$$
is just the regular Fourier transform $\mathcal F$ on the factor $\ell ^2(\Gamma)$ followed by a multiplication with
$\exp(-\I y \cdot k)$, \ie
$$
(\F \psi)(k,y)= \E^{- \I y \cdot k} (\mathcal F \otimes {\bf 1}\psi)(k,y)=
 \sum_{\gamma\in \Gamma} {\rm e}^{-i(y+\gamma)\cdot k} \psi(y+\gamma),
 $$
 for $k\in Y^*, y\in \R^d$. One immediately gets the following periodicity properties
\begin{equation}\label{perprop}
\begin{split}
(\F \psi)(k,y+\gamma) = & \ (\F \psi)(k,y),\quad \forall \gamma \in \Gamma,\\
(\F \psi)(k+\gamma^*,y) = & \ \E^{- \I y \cdot \gamma^*}(\F \psi)(k,y),\quad \forall \gamma^* \in \Gamma^*.
\end{split}
\ee
The second line in \eqref{perprop} induces a unitary representation of the group of lattice
translations in $\Gamma^*$, given by
$$
\tau: \Gamma^*\rightarrow \mathcal U(L^2(\T_y^d)),\quad \gamma^* \mapsto
\tau(\gamma^*),
$$
where $\tau(\gamma^*)$ acts as the multiplication operator by $\exp(- \I y \cdot \gamma^*)$ on
$L^2(\T_y^d)$.
Next, one easily checks that
\begin{eqnarray*}
\F(-\I \nabla_x) \F^{-1}& = & \  {\bf 1}\otimes (-\I \nabla_y) + k \otimes{\bf 1},\\
\F x \F^{-1}& = & \ \I \nabla_k^\tau.
\end{eqnarray*}
Here the operator $-\I \nabla_y$ acts on the domain $\mathcal D_1\equiv H^1(\T_y^d)$, \ie is equipped with periodic boundary conditions. On the
other hand the domain of $\I \nabla_k^\tau$ is the space of distributions in $H^1(\R^d, L^2(\T_y^d))$ which satisfy the
$y$-dependent (quasi-periodic) boundary conditions associated with the second line in \eqref{perprop}.

It is well known that the Bloch-Floquet transformation of $H(t)$, defined in \eqref{ham}, yields the fibered operator
$$
\F H(t) \F^{-1} = \int_{Y^*}^{\oplus} \D k \, H(k,t)
$$
where
\begin{equation}
H(k,t):= \frac{1}{2} (-\I \nabla_y + k)^2 + V_\Gamma(y,t),\quad k \in Y^*,
\end{equation}
with corresponding domain $\mathcal D_2\equiv H^2(\T_y^d)$, provided Assumption
\ref{asspot}. The spectrum of $H(k,t)$ is   pure point and intensively studied for
example in \cite{Wi}. The so called \emph{Bloch bands} $E_m(k,t)$, $m\in \N$, and
the \emph{Bloch projectors} $P_m(k,t)$, are consequently defined to be eigenvalues
and corresponding spectral projectors of $H(k,t)$, \ie
$$
H(k,t)P_m(k,t) = E_m(k,t) P_m(k,t),\quad m\in \N.
$$
Thereby, for definiteness, the eigenvalues $\{E_m\}_{m\in \N}$ are enumerated, according to $E_0(k,t)\leq
E_1(k,t)\leq \dots $. The corresponding (normalized)
\emph{eigenfunctions} $\{ \ph_m(k,t) \}_{m\in \N}\subset \mathcal D_2$ are called
\emph{Bloch functions}. For any fixed $k \in Y^*$, $t\in I$, they form an
orthonormal basis of $L^2(\T^d_y)$.

The \emph{extended Bloch bundle} is, by definition, the sub-bundle of the trivial bundle
\begin{equation} \label{trivial}
(Y^*_k \times \R_t)\times L^2(\T_y^d)
\end{equation}
whose fiber at the point $(k,t)$ is the range of the orthogonal
projector $P_m(k,t)$, see \cite{Pa} for a broader discussion.

\begin{remark}{\bf (Definition of the extended Bloch bundle)}
More formally, the extended Bloch bundle $\xi$ is defined in the
following way. First one introduces on the set $\R^d \times \R
\times \Hf$ the equivalence relation $\sim_{\tau}$, where
\[
(k,t,\ph) \sim_{\tau} (k', t', \ph')  \quad  \Leftrightarrow \quad
(k', t', \ph')= (k + \la \,, t,  \, \tau(\la) \ph) \quad \mbox{for
some } \la \in \Gamma^*.
\]
The equivalence class with representative $(k, t, \ph)$ is denoted
as $[k,t,\ph]$. Then the total space $E$ of the bundle $\vartheta$
is defined as
\[
E := \left \{ [k, t, \ph] \in (\R^d \times \R \times \Hf
)/{\sim_{\tau}} : \quad \ph \in \ran P_m(k,t) \right \}.
\]
This definition does not depend on the representative in view of the
covariance property (\ref{perprop}). The base space is the cylinder
$\Base = \T^* \times \R$, where $\T^* := \R^d / \Gamma^*$, and the
projection to the base space $\pi: E \to \Base $ is  $\pi[k, t, \ph]
= (\mu(k),t)$, where $\mu$ is the projection modulo  $\Gamma^*$,
$\mu: \R^d \to \T^{*}$. One checks that $\xi = (E
\stackrel{\pi}{\rightarrow} \Base)$ is a smooth complex line bundle.

Clearly, if one considers the projector $P(k,t)$ corresponding to a
family of Bloch bands, the same procedure define a complex vector
bundle over $\Base$, with typical fiber $\C^r$, $r = \dim \ran
P(k,t)$. We will use the same notation and terminology for the two
previous cases, the difference being clear from the context.
\end{remark}

In Bloch-Floquet representation the projector $P(t)={\bf 1}_{(-\infty, E(t)]}H(t)$
is again a fibered operator, in the following denoted by $P(k,t)$, and $\ran P(k,t)$
has constant dimension $M\in \N$. Assumption \ref{assgap} then implies that the
lowest $M$ Bloch bands are separated from the other bands by a finite gap. We
therefore call them \emph{isolated}.
\begin{remark} In terms of Bloch functions we have
$$
P(k,t)=\sum_{m=0}^{M} |\ph_m (k,t)\rangle \langle \ph_m (k,t)|.
$$
However, whereas $\ph_m(k,t)$ may not be a smooth function of $k$ in general, due to band crossings,
the operator $P(k,t)$ indeed is a smooth functions of $k$ due to Assumption \ref{assgap}.
\end{remark}

Finally let us state the following auxiliary result, to be used later on.

\begin{lemma}\label{trace} Let $A$ be a bounded operator acting on $L^2(\R^d)$ which is fibered in Bloch-Floquet representation, i.e.\
\[
\F A \F^{-1} =\int^\oplus_{\T^*}\D k \,A(k)\,.
\]
If, in addition, $A(k)\in \B(L^2(\T_y^d))$ is trace class, with $\tr |A(k)| < C$, for all $k\in\T^*$, then the trace per unit cell
of $A$ exists and is given by
\begin{equation}\label{traceformula}
\Tr ( A\,{\bf 1}_{Y_\lambda}) =  \frac{1}{|Y^*|} \int_{\T^*}\D k\, \tr A(k)\,,
\end{equation}
where $Y_\lambda\subset\R^d$ with $\lambda \in \R^d$ denotes any
translate of the fundamental domain $Y$ of the lattice $\Gamma$ and
${\bf 1}_{Y_\lambda}(x)$ is the characteristic function on
$Y_\lambda$. Moreover
\[
\mathcal T (A) :=  \lim_{n\to\infty}\frac{1}{|\Lambda_n|} \re \Tr ( A\,{\bf 1}_{\Lambda_n}) =
\frac{1}{(2\pi)^{d}} \int_{\T^*}\D k\,\re \tr A(k)\,.
\]
\end{lemma}

\begin{proof}
For $\lambda\in\Gamma$, let $Y_\lambda = Y+\lambda$ be an arbitrary, but fixed, $\Gamma$-translate of the unit cell $Y$ of $\Gamma$.
Having in mind that $A$ is fibered, thus $A = {{\mathscr T}_{\lambda}}^* \, A \, {\mathscr T}_{\lambda}$, with $\mathscr T_{\lambda}$ denoting the lattice
translation by $\lambda \in \Gamma$, we immediately get that
$$
\Tr(A {\bf 1}_{Y}) = \Tr( A \, \mathscr T_{\lambda} \, {\bf 1}_{Y} \, {{\mathscr T}_{\lambda}}^*) = \Tr(A \, {\bf1}_{Y_{\lambda}}).
$$
In order to evaluate the trace $ \Tr ( A\,{\bf 1}_{Y})$, we define the following orthonormal basis of $\ran {\bf 1}_{Y}\subset L^2(\R^d)$. Let
$$
g_{\gamma^*}(x) := {\bf 1}_{Y}(x)\,\E^{\I \gamma^*\cdot x }\quad \mbox{for} \
\gamma^*\in\Gamma^*\,,
$$
then
$$
\F g_{\gamma^*}(k,y) =\E^{-\I y\cdot k} \,\, \E^{\I \gamma^*\cdot y }= e_{\gamma^*}(y)\,,
$$
where for fixed $k\in\T^*$ the family of functions $e_{\gamma^*}(y) := \E^{\I(\gamma^*-k)\cdot y}$,
$\gamma^*\in\Gamma^*$, form an orthonormal basis of $L^2(\T^d)$. Hence,
\begin{eqnarray*}
 \Tr ( A\,{\bf 1}_{Y})& =& \sum_{\gamma^*} \langle g_{\gamma^*},\,  A \, {\bf 1}_{Y}\,g_{\gamma^*}\rangle
 \\&=&\frac{1}{|Y^*|}
\int_{\T^*}\D k  \sum_{\gamma^*} \int_\T\D y \,e^*_{\gamma^*}(y)\,A(k)\, e_{\gamma^*}(y) \\
&=&\frac{1}{|Y^*|} \int_{\T^*}\D k  \sum_{\gamma^*} \< e_{\gamma^*}, \, A(k) e_{\gamma^*} \>_{\Hf} =\frac{1}{|Y^*|} \int_{\T^*}\D k \, \tr
A(k)\,.
\end{eqnarray*}
For an arbitrary translate $Y+\alpha$ of $Y$, $\alpha \in \R^d$,  exactly the same
argument works with the translated basis $g_{\gamma^*}^\alpha (x):=
g_{\gamma^*}(x-\alpha)$ resp.\ $\F g^\alpha_{\gamma^*}(k,y) = \E^{\I \alpha \cdot k}
\,\, e_{\gamma^*}(y-\alpha)$.

Consequently, for an arbitrary measurable subset $\Lambda\subset Y$
the same computation with ${\bf 1}_Y$ replaced by ${\bf 1}_\Lambda$ shows that
\[
| \Tr ( A\,{\bf 1}_{\Lambda})|\leq \frac{1}{|Y^*|} \int_{\T^*}\D k \, \tr
|A(k)|\leq C\,.
\]

From this it follows that for any sequence $(\Lambda_n)$ of boxes with $\Lambda_n
\nearrow \R^d$ one has
\begin{equation}\label{TDlimit}
\mathcal T(A):=\lim_{n\to \infty}\frac{1}{|\Lambda_n|} \re \Tr (A{\bf 1}_{\Lambda_n}) = \frac{1}{|Y||Y^*|}\int_{\T^*}\D k \re \tr A(k)\,.
\end{equation}
 \end{proof}

\begin{remark} As far as the sequence $\Lambda_n \nearrow \R^d$ is concerned, one can replace the sequence of
finite volume boxes introduced for the definition of the trace per unit volume by
any \emph{F{\o}lner sequence}, \ie any sequence of measurable sets whose union adds
up to $\R^d$ and such that for any $a \in \R^d$ one has $\lim_{n \to \infty} |
(\Lambda_n +a )\setminus \Lambda_n|/ |\Lambda_n| = 0$.
\end{remark}

\subsection{The concept of equivariance}

In order to make precise statements, we have to introduce some more notations.
Let $\Hi_1$ and $\Hi_2$ be separable Hilbert spaces and $\tau_1$ and $\tau_2$ be unitary representations of
$\Gamma^*$ on $\Hi_1$ resp.\ $\Hi_2$.

We say that a function $f\in C(\R^d_k,\Hi_1)$ is $\tau_1$\emph{-equivariant}, if
\begin{equation}\label{equiv0}
f(k-\gamma^*)= \tau_1(\gamma^*)f(k)\quad \forall \, \gamma^* \in \Gamma^*, k\in \R^d.
\end{equation}

We say that a bounded operator valued function $f\in C( \R^d_k, \B(\Hi_1,\Hi_2))$ is
$(\tau_1,\tau_2)$\emph{-equivariant}, if
\begin{equation}
\label{equiv}
f(k-\gamma^*) = \tau_2(\gamma^*)\,f(k)\,\tau_1(\gamma^*)^{-1}\quad \forall \, \gamma^* \in \Gamma^*, k\in \R^d.
\end{equation}
The space of smooth $\tau_1$- resp.\ $(\tau_1,\tau_2)$-equivariant functions is then denoted by
\[
\Eq_{\Hi_1} = \{ f \in C^\infty( \R^d_k, \Hi_1 ):\, \mbox{relation \eqref{equiv0} holds} \, \}
\]
resp.\
\[
\Eq_{\Hi_1,\Hi_2} = \{ f \in C^\infty( \R^d_k, \B(\Hi_1,\Hi_2)):\, \mbox{relation \eqref{equiv} holds} \, \} .
\]
The family of seminorms
\begin{equation} \label{Seminorms}
{\|f\|}_\sigma := \sup_{k\in Y^*} {\|\partial^\sigma_k f(k)\|}_{\B(\Hi_1,\Hi_2)},
\quad \sigma \in \N_0^d, \ee turns $\Eq_{\Hi_1,\Hi_2} $ into a  Frechet space.
Here $\N_0 = \{0,1,2,\ldots \}$.

In the following there will be only two cases appearing: The unitary representation of $\Gamma^*$
on $L^2(\T^d)$, or on the domain $\mathcal{D}_2 = H^2(\T_y^d)$, which is given
by $\tau(\gamma^*)$, the operator of multiplication with $\exp(\I y\cdot\gamma^*)$.
On all other spaces (in particular on $\C^{|M|}$ below) we always use the trivial
representation $\tau\equiv{\bf 1}$. In the latter case equivariant functions are just periodic.
Therefore we will just say that a family of operators is \emph{equivariant}, understanding
that the representations of $\Gamma^*$ on the respective spaces are clear from the context.

Each  equivariant family $f\in C( \R^d_k, \B(\Hi_1,\Hi_2))  $ defines an associated operator in $\B(L^2(Y^*,\Hi_1), L^2(Y^*,\Hi_2))$ through
\[
(f\psi)(k) = f(k)\psi(k)\quad \forall \, k \in Y^*
\]
and the norms are related by
\[
{\|f\|}_{\B(L^2(Y^*, \Hi_1),L^2(Y^*,\Hi_2))} = \sup_{k\in Y^*} {\| f(k)\|}_{\B(
\Hi_1,\Hi_2)}\,.
\]
An operator in $\B(L^2(Y^*, \Hi_1),L^2(Y^*,\Hi_2))=\B(L^2(Y^*)\otimes \Hi_1,
L^2(Y^*)\otimes\Hi_2)$ is called equivariant, if it is given by an equivariant
family. Note that the composition of two equivariant operators is equivariant.

We will say that $f^\epsi\in \Eq_{\Hi_1,\Hi_2} $ \emph{is of order} $\epsi^n$
\emph{in} $\Eq_{\Hi_1,\Hi_2} $, shortly ${\| f^\epsi \|}_{\Eq_{\Hi_1,\Hi_2}}  =
\Or(\epsi^n)$, or $f^\epsi =\Or_\Eq(\epsi^n)$ and thereby abusing the norm-symbol,
if for each multi-index $\sigma\in \N_0^d$ there is a constant $C_\sigma<\infty$ such
that
\[
{\|f\|}_\sigma<C_\sigma \epsi^n\,.
\]
 In the same spirit, we say that a map $t \mapsto f(t)$ from $I \subset \R$ to $\Eq_{\Hi_1,\Hi_2}$ is differentiable,
if it is differentiable with respect to all the seminorms ${\|\cdot\|}_\sigma$ defined by \eqref{Seminorms}.



\begin{proposition}\label{propequ}
Assume that the family of bands $\{E_m(k,t)\}_{m\in M}$,
$M$ some finite index set, is isolated from the rest of the spectrum for
$(k,t)\in Y^*\times I$ and denote the corresponding spectral projection by $P(k,t)$.
Then
\[
(H(\cdot,t)-\I)^{-1}\in \Eq_{L^2(\T^d),\mathcal{D}_2}\quad\mbox{and}\quad P(\cdot,t)\in \Eq_{L^2(\T^d),\mathcal{D}_2}\,,
\quad \forall \, t\in I.
\]
Moreover if $(H(k,\cdot) -\I)^{-1}\in C^{N+1}(I, \B(L^2(\T^d),\mathcal{D}_2))$ for all $k\in Y^*$
then
$$
(H(\cdot,\cdot) -\I)^{-1}\in C^{N+1}(I,\Eq_{L^2(\T^d),\mathcal{D}_2})\quad\mbox{and}\quad P(\cdot,\cdot)\in C^{N+1}(I,\Eq_{L^2(\T^d),\mathcal{D}_2}).
$$
\end{proposition}
\begin{proof}
Let $C(t)\subset \varrho\,(H(k,t))\subset\C$ be a cycle encircling
$\{E_m(k,t)\}_{m\in M}$ once in the positive sense, but no other part of the
spectrum of $H(k,t)$. Then
\[
P(k,t) = \frac{\I}{2\pi}\oint_{C(t)}\D z\,(H(k,t)-z)^{-1}\,.
\]
Then the statements about $P(k,t)$ follow from the corresponding statements about the   resolvent.
 For $z\in\varrho \,(H(k,t))$ we have
\begin{eqnarray*}
R_z(k-\gamma^*,t) &= & (\tau(\gamma^*)(H(k,t)-z)\tau(\gamma^*)^{-1})^{-1}\\
&=& \tau(\gamma^*) (H(k,t)-z)^{-1}\tau(\gamma^*)^{-1}\\
&=& \tau(\gamma^*) R_z(k,t) \tau(\gamma^*)^{-1}\,,
\end{eqnarray*}
and thus $R_z(\cdot,t)$ is equivariant.
The statements about the differentiability  follow from
\[
\partial_{k_j}R_z(k,t) = - R_z(k,t)\,(\partial_{k_j}H(k,t))\,R_z(k,t)
\]
resp.\
\[
\partial_{t}R_z(k,t) = - R_z(k,t)\,(\partial_{t}H(k,t))\,R_z(k,t)
\]
and iterations of these formulas.
\end{proof}

For the proof of the super-adiabatic theorem we shall also need the following observation:
\begin{lemma}\label{basis}
There exists a smooth $\tau$-equivariant orthonormal basis $\left(\chi_\alpha
(k,t)\right)_{\alpha=1}^M$ of $ \ran P(k,t)$ such that the coefficient of the Berry
connection in the time direction vanishes identically in this basis. More precisely,
for all $(t,k)\in I\times Y^*$ and $\alpha,\beta\in \{1,\ldots,M\}$ it holds that
\begin{equation*}\label{berry}
\phi_{\alpha\beta}(k,t) := \I \langle \chi_\alpha(k,t),\partial_t \chi_\beta(k,t)\rangle_{L^2(\T^d_y)} \equiv 0\,.
\end{equation*}

If $P(\cdot,\cdot)\in C^{N+1}(I,\Eq_{L^2(\T^d),\mathcal{D}_2})$, then $\chi_\alpha(\cdot,\cdot)\in C^{N+1}(I,\Eq_{\mathcal{D}_2})$.
\end{lemma}
\begin{proof}
First, we choose at time $t=0$, a smooth and $\tau$-equivariant orthonormal basis $\chi(k,0) =
\left(\chi_1 (k,0),\ldots,\chi_{M}(k,0)\right)$ of $\ran P(k,0)$. We can always find such a basis since the
complex vector bundle (over the torus) defined by $P(k,0)$ is trivial. As shown in \cite{Pa}, triviality of the Bloch bundle with $M$-dimensional fiber is a consequence of the
time-reversal symmetry of the Hamiltonian \eqref{ham} also for $M>1$. Then we determine $\chi(k,t) =
\left(\chi_1 (k,t),\ldots,\chi_{M}(k,t)\right)$ as the solution of the equation
\[
\partial_t \chi_\alpha(k,t) = [\partial_t P(k,t), P(k,t)] \,\chi_\alpha(k,t)\,.
\]
According to Kato \cite{Ka} $\chi(k,t) = \left(\chi_1 (k,t),\ldots,\chi_{M}(k,t)\right)$ is an orthonormal basis
of $\ran P(k,t)$ and satisfies
\[
\langle \chi_\beta(k,t) ,\partial_t \chi_\alpha(k,t)\rangle = \langle \chi_\beta(k,t) , [\partial_t P(k,t), P(k,t)]  \chi_\alpha(k,t)\rangle = 0,
\]
since
$$
\partial_t P(k,t) =   P(k,t)\partial_t P(k,t) P(k,t)^\perp + P(k,t)^\perp \partial_t P(k,t)P(k,t).
$$
Also, since $P(k,t)$ and $\partial_t P(k,t)$ are equivariant, so are the $\chi_\alpha(k,t)$'s.
\end{proof}
Note that, due to possible band crossings within $\ran P(k,0)$, the initially chosen smooth basis elements
$\left(\chi_\alpha (k,0)\right)_{\alpha=1}^M$ in general
are \emph{not} Bloch functions, \ie they are not eigenfunctions  of the operator $H(k,t)$.


\begin{remark} In the case where $M =1$, the choice
$$
\chi(k,t) = \E^{- \I \int_0^T \phi(s) \D s} \chi(k,0),
$$
has the desired property stated in the lemma above.
\end{remark}

\section{The Super-adiabatic theorem}\label{super}

To perform an approximation of \eqref{model1} to sufficient high order in $\e$,
we need the following  so called \emph{super-adiabatic} theorem.


 Recall the notation $\mathcal{D}_1= H^1(\T^d_y)$ and $\mathcal{D}_2 = H^2(\T^d_y)$.

\begin{proposition}\label{adia}
For each isolated family of Bloch bands $\{E_m(k,t)\}_{m\in M}$, with $M\subset \mathbb N$ some finite index set,
there exists an  equivariant family of  projections $P_N^\epsi(k,t)\in C^{ 1}(I,\Eq_{L^2(\T^d),\mathcal{D}_2})$ and an  equivariant family of unitary operators
$$
T^\epsi_N(k,t): P^\epsi_N(k,t)\,L^2(\T^d) \to \C^{|M|}
$$
such that $P_N^\epsi(k,0)=P(k,0)$, $P_N^\epsi(k,T)=P(k,T)$ and
the  following statements hold:
\begin{enumerate}
\item[\textbf{(A)}] {\bf Adiabatic decoupling:} The
propagator $U^\epsi(k,t)$ (recall that $t_0=0$) of the Schr\"odinger equation
restricted to $\ran P_N^\epsi(k,0)=$ $\ran P(k,0)$ is close to the adiabatic propagator $U_{\rm a}^\epsi(k,t)$ generated by the adiabatic Hamiltonian
$$
H_{\rm a}^\epsi (k,t) = P_N^\epsi(k,t) H(k,t) P_N^\epsi(k,t) + \I\epsi[\partial_t P_N^\epsi(k,t), P_N^\epsi(k,t)]
$$
up to errors of order $\epsi^N$. More precisely we have
\begin{equation}
\label{adiStat0}
{\|\, \left(U^\epsi(k,t)-U^\epsi_{\rm a}(k,t)\right)\,P_N^\epsi(k,0) \|}_{\B(L^2(\T^d),\mathcal{D}_1)} = \Or(\epsi^N)\,
\end{equation}
and
\begin{equation}\label{adiStat1}
{\|\, \left(U^\epsi(k,t)-U^\epsi_{\rm a}(k,t)\right)\,P_N^\epsi(k,0) \|}_{\B(L^2(\T^d),L^2(\T^d))} = \Or(\epsi^N|t|)\,.
\end{equation}
Since, by construction,   $U^\epsi_{\rm a}(k,t) P_N^\epsi(k,0) = P_N^\epsi(k,t) U^\epsi_{\rm a}(k,t) $,
it follows that \emph{Ran}$P_N^\epsi(k,t)$ is almost invariant under the true time-evolution, i.e.
\begin{equation}\label{adiStat}
{\|\,({\bf 1} - P_N^\epsi(k,t))\,U^\epsi(k,t)\,P_N^\epsi(k,0)\,\|}_{\B(L^2(\T^d),\mathcal{D}_1)} = \Or(\epsi^N)\,
\end{equation}
and that $P^\epsi_N(k,t)$ approximates $\rho^\e(k,t) = U^\e(k,t)\,P(k,0)\,U^\e(k,t)^*$,
\begin{equation}\label{adiStat3}
{\|\,P^\e_N(k,t)- \rho^\e(k,t)\,  \|}_{\B(L^2(\T^d),\mathcal{D}_1)} = \Or(\epsi^N)\,.
\end{equation}
\item[\textbf{(B)}]{\bf Effective dynamics:}
Let the effective propagator  $U_{\rm eff}^\epsi(k,t)$ on $\C^{|M|}$ be defined by
\[
U^\epsi_{\rm eff}(k,t) := T_N^\epsi (k,t) \,  U^\epsi_{\rm a}(k,t)  \, T_N^\epsi(k,0)^*\,,
\]
then $U^\epsi_{\rm eff}(k,t)$ solves  the effective Schr\"odinger equation
\[
\I\,\epsi\frac{\D}{\D t}\,U^\epsi_{\rm eff }(k,t) = H_{\rm eff}(k,t)\,U^\epsi_{\rm eff }(k,t)\]
with periodic effective Hamiltonian
\[
H_{\rm eff}(k,t) = {\bf E}(k,t)  + \Or(\epsi^2)\,.
\]
The self-adjoint $|M|\times|M|$-matrix ${\bf E}(k,t)$ is $\Gamma^*$-periodic in $k$ and given by
\[
{\bf E}_{\alpha\beta} (k,t)= \langle \chi_\alpha(k,t), \,H(k,t)\,\chi_\beta(k,t)\rangle_{L^2(\T_y^d)}\,.
\]
Here $(\chi_\alpha(k,t))_{\alpha=1}^M$ is a basis as in Lemma~\ref{basis}, which is used to construct $T^\epsi_N(k,t)$.

\end{enumerate}
If not stated explicitly otherwise, all estimates are  uniform for $t$ in a finite interval.
\end{proposition}

 \begin{proof}
By Assumption \ref{asspot}, we have $V_\Gamma \in C^{N+1}(I,\B(H^2(\R^d),L^2(\R^d)))$, which
 implies, that $(H(\cdot)-\I)^{-1}\in C^{N+1}(I,\B(L^2(\R^d), H^2(\R^d)))$, or that
 $(H(k,\cdot) -\I)^{-1}\in C^{N+1}(I,  \B(L^2(\T^d),\mathcal{D}_2))$ fiberwise in Bloch-Floquet representation.
 Hence, according to Proposition~\ref{propequ} we have that
 \begin{equation}\label{smooth}
(H(\cdot,\cdot) -\I)^{-1}\in C^{N+1}(I,\Eq_{L^2(\T^d),\mathcal{D}_2})\ \mbox{and}\ P(\cdot,\cdot)\in C^{N+1}(I,\Eq_{L^2(\T^d),\mathcal{D}_2})\,.
\end{equation}

 The projector  $P_N^\epsi(k,t)$ for {\em fixed} $k$ is constructed using Nenciu's scheme  \cite{Ne}.
 For convenience of the reader and in order to observe equivariance we briefly state the result:

 Denote the resolvent of $H(k,t)$ at the point $z\in \C$ by $R_z(k,t)=(H(k,t)-z)^{-1}$
 and define $P_j(k,t)$, for $j\leq N$, recursively via
 \begin{eqnarray*}
 P_{j}(k,t) &:= &\  G_j(k,t)-2P_0(k,t)G_j(k,t)P_0(k,t) \\
 && +\,\frac{1}{2\pi}\oint_{C(t)}\D z\, R_z(k,t)\left[ P_0(k,t), \partial_t P_{j-1}(k,t)\right]\,R_z(k,t)\,,
 \end{eqnarray*}
 where $C(t)\subset\varrho \,(H(k,t))\subset\C$ is a cycle encircling $\{E_m(k,t)\}_{m\in M}$ once in the positive sense,
 but no other part of the spectrum of $H(k,t)$, $P_0(k,t):= P(k,t)$ and
 \[
 G_j(k,t) := \sum_{m=1}^{j-1}P_m(k,t)P_{j-m}(k,t)\,.
 \]
 Then
 \[
 \widetilde P_N^\epsi(k,t) := \sum_{j=0}^N \epsi^j P_j(k,t)
 \]
  satisfies
 \begin{equation}\label{almostP}
( \widetilde P_N^\epsi(k,t)\,)^2-\widetilde P_N^\epsi(k,t) = \epsi^{N+1} G_{N+1}(k,t)
 \end{equation}
 as well as
 \begin{equation}\label{almostK}
 \big[ \I\epsi\partial_t - H(k,t),\,\widetilde P_N^\epsi(k,t)\big] = \epsi^{N+1}\partial_t P_N(k,t)\,.
 \end{equation}
 From \eqref{smooth} and the fact that compositions of equivariant operators are equivariant it follows that $P_j(\cdot,\cdot) \in C^{N+1-j}(I,\Eq_{L^2(\T^d),\mathcal{D}_2})$ for $j=1\ldots,N$, and thus  $  \widetilde P_N^\epsi(\cdot,\cdot)\in C^{ 1}(I,\Eq_{L^2(\T^d),\mathcal{D}_2})$
 and also $G_{N+1}(\cdot,\cdot)\in C^{ 1}(I,\Eq_{L^2(\T^d),\mathcal{D}_2})$.

According to \eqref{almostP} the spectrum of $\widetilde P^\epsi_N(k,t)$ is located in  $\epsi^{N+1}$ neighborhoods of $0$ and $1$. Thus for $\epsi$ sufficiently small one can
 define $P_N^\epsi(k,t)$ as the spectral projection of $\widetilde P_N^\epsi(k,t)$ associated with its spectrum near $1$, \ie
 \[
 P^\epsi_N(k,t) = \frac{\I}{2\pi}\oint_{|z-1|=\frac{1}{2}}\D z\,(\widetilde P^\epsi_N(k,t)-z)^{-1}\,.
 \]
 By this definition  $P_N^\epsi(k,t)$ is obviously a linear operator from $L^2(\T^d)$ into $\mathcal{D}_2$ and an element of $\Eq_{L^2(\T^d),L^2(\T^d)} \cap \Eq_{\mathcal{D}_2,\mathcal{D}_2} $.
Since $\|P^\epsi_N(k,t)-\widetilde P^\epsi_N(k,t)\| = \Or(\epsi^{N+1})$, it follows that $\|P^\epsi_N(k,t)-P_0(k,t)\| = \Or(\epsi)$ and hence
dim($\ran  P^\epsi_N(k,t))$ = dim($\ran P_0(k,t))=|M|<\infty$. Thus $P_N^\epsi(k,t)$ and its derivatives have finite rank, which implies
$P_N^\epsi(k,t)\in C^{ 1}(I,\Eq_{L^2(\T^d),\mathcal{D}_2})$.

From \eqref{almostK} we obtain
 \begin{eqnarray*}\lefteqn{
  \ \big[ \I\epsi\partial_t - H(k,t),\, P_N^\epsi(k,t)\big] =}\\&&
 = -\,  \frac{\I \epsi^{N+1}}{2\pi}\oint_{|z-1|=\frac{1}{2}}\D z \,
  (\widetilde P^\epsi_N(k,t)-z)^{-1} \, \partial_t P_N(k,t) \, (\widetilde P^\epsi_N(k,t)-z)^{-1}.
 \end{eqnarray*}
 Hence
 \begin{equation}\label{dtHcom}
 {\left \|\, \big[ \I\epsi\partial_t - H(k,t),\, P_N^\epsi(k,t)\big]\,\right \|}_{\Eq_{L^2(\T^d),\mathcal{D}_2}}=\Or(\epsi^{N+1})
 \end{equation}
 and also
 \begin{equation}\label{HCommu}
 {\left \|\, \big[  \I\epsi\partial_t - H(k,t),\, P_N^\epsi(k,t)\big]\,\right \|}_{\Eq_{\mathcal{D}_1}}=\Or(\epsi^{N+1}).
 \end{equation}
 Now \eqref{adiStat} follows from Kato's construction \cite{Ka}. Indeed, let
 \[
 H^\e_{\rm a}(k,t) = P^\epsi_N(k,t) H(k,t) P^\epsi_N(k,t) + \I\epsi [ P^\epsi_N(k,t), \partial_t  P^\epsi_N(k,t)]
 \]
 be the adiabatic Hamiltonian and $U^\epsi_{\rm a}(k,t)$ the adiabatic evolution generated by $H^\epsi_{\rm a}(k,t)$. Then by construction one has
 \begin{equation}\label{Kato}
 U^\epsi_{\rm a}(k,t) P^\epsi_N(k,0) = P^\epsi_N(k,t)U^\epsi_{\rm a}(k,t)\,.
 \end{equation}
Clearly \eqref{Kato} holds for $t=0$ and multiplying both sides by  $U^\epsi_{\rm a}(k,t)^*$ and differentiating with respect to $t$ shows that the equality holds for all times. As a consequence we find that
 \begin{align*}
 &\ (U^\epsi(k,t)-U^\epsi_{\rm a}(k,t))P^\epsi_N(k,0) = \\
 &= -U^\epsi(k,t)\int_0^t\D s\, \frac{\D}{\D s}\left( U^\epsi(k,-s) U^\epsi_{\rm a}(k,s) \right) P^\epsi_N(k,0)\\
 &= -\frac{\I}{\epsi}\, U^\epsi(k,t)\int_0^t\D s\,
 U^\epsi(k,-s) \left( H^\epsi(k,s) - H^\epsi_{\rm a}(k,s)\right) P^\epsi_N(k,s) U^\epsi_{\rm a}(k,s)\\
 &= \frac{\I}{\epsi} \, U^\epsi(k,t)\int_0^t\D s\, U^\epsi(k,-s) \big[\I\epsi\partial_s - H^\epsi(k,s), P^\epsi_N(k,s)\big] P^\epsi_N(k,s)  U^\epsi_{\rm a}(k,s) \\
 &= \frac{\I}{\epsi} \, U^\epsi(k,t)\int_0^t\D s\, U^\epsi(k,-s) \big[\I\epsi\partial_s - H^\epsi(k,s), P^\epsi_N(k,s)\big] U^\epsi_{\rm a}(k,s) P^\epsi_N(k,0)  \,.
 \end{align*}
  As to be shown in Lemma~\ref{BLemma} below, $U_{\rm a}^\epsi(k,t)$ and $U^\epsi(k,t)$
 are  bounded operators  from $\mathcal{D}_1$ to $\mathcal{D}_1$.
 Hence the previous computation
 together with \eqref{HCommu} yields \eqref{adiStat0}.
 The statement \eqref{adiStat1} then follows analogously, but
 since we do not need to invoke Lemma~\ref{BLemma} below, we additionally obtain an error estimate linear in $t$.
 Thus for the full time-evolution we find \eqref{adiStat},
  \begin{equation*}\begin{split}
& {\big \|\,({\bf 1} - P_N^\epsi(k,t))\,U^\epsi(k,t)\,P_N^\epsi(k,0)\big\|}_{\B(L^2(\T^d),\mathcal{D}_1) }=\\
& = {\big \|\,({\bf 1} - P_N^\epsi(k,t))\,\left(U^\epsi(k,t)-U^\epsi_{\rm a}(k,t)\right)\,P_N^\epsi(k,0)\big\|}_{\B(L^2(\T^d),\mathcal{D}_1)} \\
&\leq \  {\big \|\,\left(U^\epsi(k,t)-U^\epsi_{\rm a}(k,t)\right)\,P_N^\epsi(k,0)\big\|}_{\B(L^2(\T^d),\mathcal{D}_1)}=  \Or(\epsi^N)\,.
\end{split}
   \end{equation*}
  \begin{lemma}\label{BLemma}
  $U^\epsi(k,t)$, $U^\epsi_{\rm a}(k,t)  \in \B(\mathcal{D}_1)$ uniformly  for $(k,t)\in Y^*\times I$.
  \end{lemma}
 \begin{proof}

 From Assumption  \ref{asspot} and the fact the $\nabla$ is infinitesimally bounded with respect to $\Delta$ it follows that there are  constants $\mu_\pm,\nu_\pm> 0$ such that  for all $\psi\in\mathcal{D}_2$ and $t\in I$
 \[
 \mu_-\|\nabla\psi\|^2 -\nu_- \|\psi\|^2\leq \langle \psi,H(k,t)\psi \rangle\leq  \mu_+\|\nabla\psi\|^2 +\nu_+ \|\psi\|^2
\,.
 \]
 Hence
  \[
  {\|\psi\|}_t^2 := \langle\psi, H(k,t)\psi\rangle_{L^2} +(\nu_-+\mu_-) {\|\psi\|}_{L^2}^2\,,
  \]
 defines a norm on $\mathcal{D}_1=H^1(\T^d)$ which is
 equivalent to ${\|\psi\|}_{\mathcal{D}_1}^2 :=  \|\nabla\psi\|^2+\|\psi\|^2$
  uniformly for $t\in I$, since
  \[
   \mu_-{\|\psi\|}_{\mathcal{D}_1}^2\leq   {\|\psi\|}_t^2  \leq (\mu_++\mu_-+\nu_++\nu_-){\|\psi\|}_{\mathcal{D}_1}^2\,.
  \]
  For $\psi_0\in \mathcal{D}_2$ let $\psi(t) = U^\epsi(k,t)\psi_0$. Then by the assumption that $\dot V_\Gamma(y,t)$ is relatively form bounded with respect to $\Delta$ there is a constant $c>0$ such that
 \begin{align*}
  \frac{\D}{\D t}  {\|\psi(t)\|}_t^2 = & \ \langle\psi(t),\dot H(k,t)\psi(t)\rangle \\
  = &\  \langle\psi(t),\dot V_\Gamma(t)\psi(t)\rangle\leq c\mu_-{\|\psi(t)\|}^2_{\mathcal{D}_1}\leq c\,  {\|\psi(t)\|}_t^2\,.
  \end{align*}
A Gronwall lemma then yields ${\|\psi(t)\|}_t^2\leq \E^{ct} {\|\psi(0)\|}_0^2$ and thus we get that
${\|U^\epsi(k,t)\|}_{\mathcal{B}(\mathcal{D}_1)}<\infty $.
%
%
Since the generators $H(k,t)$ and $H_{\rm a}^\epsi(k,t)$ differ only by a bounded operator, the same argument shows that also $U^\epsi_{\rm a}(k,t) \in \B(\mathcal{D}_1)$.
 \end{proof}

 For the construction of the unitary $T_N^\epsi(k,t)$ we first choose an orthonormal basis
 $ \left(\chi_1 (k,t),\ldots,\chi_{|M|}(k,t)\right)$ of $\ran P(k,t)$ such that
 $\chi_\alpha(\cdot,\cdot)\in C^{N+1}(I,\Eq_{\mathcal{D}_2})$ and
$$
\phi_{\alpha\beta}(k,t) = \I \langle \chi_\alpha(k,t),\partial_t \chi_\beta(k,t)\rangle \equiv 0
$$
for all $t\in I\,,\,\,\alpha,\beta\in \{0,\ldots,|M|\}$. This is always possible as shown in Lemma~\ref{basis}.
Next we define $T_0(k,t): L^2(\T^d)\to \C^{|M|}$ through
\[
(T_0(k,t) \psi)_\alpha = {\langle \chi_\alpha(k,t),\,\psi\rangle}_{L^2(\T_y^d)}\,.
\]
Hence $T_0(\cdot,\cdot)\in C^{N+1}(I,\Eq_{L^2(\T^d),\C^{|M|}})$.
We now give a simplified version of the construction developed in \cite{PST}. The idea is again to construct first an asymptotic expansion to the appropriate order. Let
\[
\widetilde T^\epsi_n(k,t) = \sum_{j=0}^n \epsi^j T_j(k,t).
\]
We require that at order $n\in \N$ it holds that
\begin{eqnarray}
\widetilde T_n^\epsi(k,t) \,\widetilde T_n^\epsi(k,t)^*- {\bf 1}_{\C^{|M|}} &=&\Or(\epsi^{n+1})\,,\nonumber\\[-2mm]
\label{Tcond}\\[-2mm]
\widetilde T_n^\epsi(k,t)\,({\bf 1}-P_N^\epsi(k,t) )&=&\Or(\epsi^{n+1})\,,\nonumber
\end{eqnarray}
\ie that $T_n^\epsi(k,t)^*$ is almost unitary as a map from $\C^{|M|}$ to its range and
that the range is almost that of $P_N^\epsi(k,t)$.  Making the ansatz
\[
T_j(k,t) = T_0(k,t) \left( a_j(k,t) + b_j(k,t) \right),
\]
with self-adjoint $a_j(k,t)$ and anti-self-adjoint $b_j(k,t)$, we find the following recurrence for the coefficients.
Clearly \eqref{Tcond} holds for $n=0$. (For better readability, we drop the $(k,t)$-dependence in the following,
but it is understood that all operators appearing depend on $k$ and $t$ and are equivariant.)
Assume that \eqref{Tcond} holds for $n$, then
\[
\widetilde T_n^\epsi  \,\widetilde T_n^{\epsi *}- {\bf 1}_{\C^{|M|}} =\epsi^{n+1}\sum_{k=1}^{n}T_k T_{n+1-k}^*
+\Or(\epsi^{n+2}) =:\epsi^{n+1}\,A_{n+1} +\Or(\epsi^{n+2})
\]
and thus $T_{n+1} $ has to solve
\begin{equation}\label{rela}
T_0 T_{n+1}^* + T_{n+1} T_{0} ^* = 2 T_0\, a_{n+1}\, T^*_0  \stackrel{!}{=} - A_{n+1} \,.
\end{equation}
Hence, one must choose
$$
 a_{n+1} = -{\textstyle \frac{1}{2}}T_0^*A_{n+1}T_0\,.
$$
Again by the induction assumption
\begin{eqnarray*}\lefteqn{
(\widetilde T_n^\epsi + \epsi^{n+1}T_0 a_{n+1})({\bf 1}- P_N^\epsi)=}\\ &= & - \epsi^{n+1} \sum_{j=0}^n \left(T_j P_{n+1-j}
 -{\textstyle \frac{1}{2}} A_{n+1}T_0({\bf 1}-P_0)\right) + \Or(\epsi^{n+2})\\
&=&  -\epsi^{n+1}   \sum_{j=0}^n T_j P_{n+1-j}  + \Or(\epsi^{n+2})\\
& =:  &\epsi^{n+1} B_{n+1} + \Or(\epsi^{n+2})\,.
\end{eqnarray*}
Thus $b_{n+1}$ has to solve
$$
T_0 b_{n+1}({\bf 1}- P_0) = - B_{n+1}= -B_{n+1}({\bf 1}- P_0) +\Or(\epsi)\,,
$$
where the last equality follows by multiplying the previous equation with $P_N^\epsi$ from the right.
A possible choice is therefore
$$
b_{n+1} = -T_0^* B_{n+1}({\bf 1}- P_0)\,,
$$
which leaves us with the following recurrence relation
$$
T_{n+1} = -\frac{1}{2}\sum_{k=1}^{n}T_k T_{n+1-k}^*  T_0 +
 \sum_{j=0}^n T_j P_{n+1-j}  ({\bf 1}- P_0),
$$
up to $n+1=N$.
Then by construction $\widetilde T_N^\epsi(\cdot,\cdot) \in C^{1}(I,\Eq_{L^2(\T^d),\C^{|M|}})$ and
\begin{eqnarray*}
\widetilde T_N^\epsi(k,t) \,\widetilde
T_N^\epsi(k,t)^*- {\bf 1}_{\C^{|M|}} &=&\Or(\epsi^{N+1}),\\
\widetilde T_N^\epsi(k,t)\,({\bf 1}-P_N^\epsi(k,t) )&=&\Or(\epsi^{N+1})\,.
\end{eqnarray*}
In order to construct a true unitary observe that $\widetilde T^\epsi_N\,\widetilde T^{\epsi\,*}_N$ is a positive self-adjoint operator $\Or(\epsi^{N+1})$-close to the identity. Hence
\[
\hat T_N^\epsi = \left( \widetilde T^\epsi_N\,\widetilde T^{\epsi\,*}_N \right)^{-\frac{1}{2}} \widetilde T_N^\epsi
\]
is a unitary operator  $\Or(\epsi^{N+1})$-close to $\widetilde T_N^\epsi$. Finally let
 \[
T^\epsi_N =        \left(    \hat T_N^\epsi \,P_N^\epsi\,\hat T^{\epsi\,*}_N\right)^{-\frac{1}{2}}      \, \hat  T_N^\epsi \,P_N^\epsi\,,
\]
which implies $T_N^\epsi\,T_N^{\epsi\,*} = P_N^\epsi$ and again $T_N^\epsi(\cdot,\cdot) \in C^{1}(I,\Eq_{L^2(\T^d),\C^{|M|}})$

We conclude that $P_N^\epsi(k,t)$ and $T_N^\epsi(k,t)$ are equivariant by construction
and have an explicit expansion up to terms of order $\Or(\epsi^{N+1})$ given by
\[
P_N^\epsi(k,t) = \sum_{j=0}^N \epsi^j P_j(k,t) + R_{P,N+1}^\epsi(k,t)
\]
and
\[
T_N^\epsi(k,t) = \sum_{j=0}^N \epsi^j T_j(k,t) + R_{T,N+1}^\epsi(k,t)\,,
\]
where
\[
{\|R_{P,N+1}^\epsi(k,t)\|}_\Eq=\Or(\epsi^{N+1})\quad\mbox{and}\quad {\|R_{T,N+1}^\epsi(k,t)\|}_\Eq=\Or(\epsi^{N+1})\,.
\]
In particular we have
\[
P_1(k,t) = \frac{1}{2\pi}\oint_{C(t)}\D z\, R_z(k,t)\left[ P_0(k,t), \partial_t P_0(k,t)\right]\,R_z(k,t)
\]
and
\[
T_1(k,t) = T_0(k,t) \,P_1(k,t)\,({\bf 1}-P_0(k,t))\,,
\]
which yields $T_1P_0 = T_0P_1P_0 =0$.
Now
\begin{align*}
H_{\rm eff}^\epsi(k,t) := & \  T^\epsi_N(k,t)\, P_N^\epsi(k,t)\,H(k,t)\,P_N^\epsi(k,t)\,T_N^{\epsi}(k,t)^*\\
& \, - \I \e \,  T_N^{\epsi}(k,t) \, \partial_t T_N^{\epsi}(k,t)^*
\end{align*}
defines a self-adjoint and periodic operator on $\C^{|M|}$, which depends continuously on $t$. More precisely, $H_{\rm eff}^\epsi(\cdot,\cdot)\in C(I,\Eq_{\C^{|M|},\C^{|M|}})$.
Due to the particular choice of our basis $\left(\chi_\alpha (k,t)\right)_{\alpha=1}^M$,
constructed in Lemma~\ref{basis}, we get that
\begin{align*}
 T_N^{\epsi}(k,t) \, \partial_t T_N^{\epsi}(k,t)^* = & \
 \sum_\alpha |e_\alpha\rangle \langle\chi_\alpha(k,t)| \sum_\beta| \partial_t \chi_\beta(k,t)\rangle\langle e_\beta|+\O_\Eq(\e)\\
 = & \ \O_\Eq(\e).
\end{align*}
Therefore, $T_1P_0 = T_0P_1P_0 =0$ implies
\begin{align*}
H_{\rm eff}^\epsi(k,t) = & \ T_0(k,t) P_0(k,t) H(k,t) P_0(k,t) T_0(k,t)^* +\Or_\Eq(\epsi^2)\\
= & \ {\bf E}(k,t) + \Or_\Eq(\epsi^2)
\,.
\end{align*}
Since
\[
T^\epsi_N(k,t)\left( \I\frac{\D}{\D t} - H_{\rm a}^\epsi(k,t)\right) T^\epsi_N(k,t)^* =  \I\frac{\D}{\D t} - H_{\rm eff}^\epsi(k,t)\,,
\]
it follows that
\begin{equation}\label{intert}
U^\epsi_{\rm a}(k,t) \,T^{\epsi\,*}_N(k,0) = T^{\epsi\,*}_N(k,t) U^\epsi_{\rm eff}(k,t)\,,
\end{equation}
which concludes the proof.
\end{proof}


\section{Proof of the main results}\label{sproof}


While Theorem~\ref{th1} follows directly from the super-adiabatic approximation \eqref{adiStat3} of the state at time $t$, the semiclassical approximation of Theorem~\ref{th2} is based on the effective dynamics in the almost invariant subspace.

\subsection{Proof of Theorem~\ref{th1}}

We first note that the state
\[
\rho^\epsi(k,t)= U^\epsi(k,t)P(k,0)U^\epsi(k,t)^*
\]
 at time $t\in I$
as well as the current operator $J^\e$ defined in \eqref{curr} are fibered,
\begin{equation*}
 \F J^\e \F^{-1}(k) =\frac{1}{\epsi}\, (-\I\nabla_y+k) = \frac{\I}{\epsi} \, [H(k,t),\I\nabla_k^\tau]\,.
\end{equation*}
Since $\rho^\epsi(k,t)$ has finite dimensional range contained in
$\mathcal{D}_1$, $\rho^\e(k,t)\,J^\epsi(k)$ is trace class and,
invoking Lemma~\ref{trace}, the macroscopic current in the state
$\rho^\e(t)$ is  given by
\begin{equation*}
\dot\p^\e(t)=  \frac{1}{(2\pi)^d}\int_{\T^*}\D k\, \re \tr \rho^\epsi(k,t)J^\e(k)
= \frac{1}{(2\pi)^d}\int_{\T^*}\D k\, \tr \rho^\e(k,t)J^\e(k)  \,.
\end{equation*}
The  integrand can be evaluated as follows:
First observe that with \eqref{adiStat3}
\begin{equation*}
\begin{split}
 \tr \rho^\e(k,t)\,J^\e(k) = &\, \tr P^\e_N(k,t)\,J^\e(k) + \Or(\e^N) \\
 = & \,  {\frac{\I}{\epsi} }\tr  P^\e_N(k,t)[H(k,t), \I\nabla_k^\tau] P^\e_N(k,t)  +\Or(\epsi^N).
\end{split}
\end{equation*}
Next we compute using \eqref{dtHcom} that
\begin{align*}
P^\epsi_N\,H\, \I\nabla_k^\tau\, P^\e_N
= & \ P^\e_N \,H\, P^\e_N \, \I\nabla_k^\tau  +\I P^\epsi_N\,H\, ( \nabla_k P^\e_N)\,( P^\e_N + {\bf 1}-P^\e_N)\\
 =& \  P^\e_N \,H\, P^\e_N \, \I\nabla_k^\tau\ - \I P^\epsi_N\,[H, P_N^\e]\, ( \nabla_k P^\e_N)\, P^\e_N\\
 & \ +\I P^\epsi_N\,H\, ( \nabla_k P^\e_N)\,( {\bf 1}-P^\e_N)\\
  =&\  P^\e_N \,H\, P^\e_N \, \I\nabla_k^\tau  + \e\, P^\epsi_N\, \dot P_N^\e\, ( \nabla_k P^\e_N)\, P^\e_N\\
  &\ +\I P^\epsi_N\,H\, ( \nabla_k P^\e_N)\,( {\bf 1}-P^\e_N)  + \Or_\Eq(\epsi^{N+1})
\end{align*}
and therefore
\begin{align*}
P^\epsi_N\,[H,\, \I\nabla_k^\tau]\,P^{\epsi}_N =& \, P^\epsi_N\,H\, \I\nabla_k^\tau\,P^{\epsi}_N-(P^\epsi_N\,H\, \I\nabla_k^\tau\,P^{\epsi}_N)^*\\
 = & \,
[P^\epsi_N\,H\, P^{\epsi\,*}_N,\, \I\nabla_k^\tau\, ] + \epsi \,P^\epsi_N\,[\dot P^\epsi_N,\,(\nabla_k P^\epsi_N)]\,P^{\epsi}_N \\
 & \, +R^\e_N +\Or_\Eq(\epsi^{N+1})\\
=& \,  -\I(\nabla_k P^\epsi_N\,H\, P^{\epsi}_N) + \epsi \,P^\epsi_N\,[\dot P^\epsi_N,\,(\nabla_k P^\epsi_N)]\,P^{\epsi}_N \\
& \, +R^\e_N +\Or_\Eq(\epsi^{N+1})\,,
\end{align*}
where $\Tr R^\e_N(k,t) = 0$ for all $t\in I$ and $k\in  \T^*$.
We take the trace  for fixed $k$ and abbreviate  ${\bf E}^\epsi(k,t)=P^\epsi_N(k,t)\,H(k,t)\, P^{\epsi}_N(k,t)$ to obtain
\begin{equation*}
\begin{split}
\tr \( P^\epsi_N(k,t)[H(k,t),  \I \nabla_k^\tau]   P^\epsi_N(k,t) \) =  &  -\I \nabla_k\tr {\bf E}^\epsi(k,t) \\
 &  + \epsi \tr \( P^\epsi_N(k,t) [\partial_t P^\epsi_N (k,t), \nabla_k P^\epsi_N(k,t)] \) \\
 & + \Or(\epsi^{N+1}).
 \end{split}
\end{equation*}
In summary we just computed that
$$
\tr \rho^\e(k,t) J^\epsi(k) ={ \frac{1}{\epsi}}\nabla_k\tr {\bf E}^\epsi(k,t) + \I \tr P^\epsi_N(k,t)
[\partial_t P^\epsi_N(k,t),\nabla_k P^\epsi_N(k,t)]  +\Or(\epsi^{N}).
$$
By exploiting the fact that $\tr{\bf E}^\epsi(k,t)$ is a periodic function of $k$,
$$
\tr{\bf E}^\epsi(k-\gamma^*,t)=\tr \tau(\gamma^*)\,{\bf E}^\epsi(k,t)\,\tau(\gamma^*)^{-1}=\tr{\bf E}^\epsi(k,t)
\quad \forall \,\gamma^*\in\Gamma^*\,,
$$
we finally get
\begin{eqnarray*}
\dot\p^\e(t)&= & \  \frac{1}{(2\pi)^d}\int_{\T^*}\D k\, \tr \( \rho^\e(k,t)J^\e(k) \) \\
&= &  \frac{\I}{(2\pi)^d} \int_{\T^*}\D k\, \tr \( P^\epsi_N(k,t) [\partial_t P^\epsi_N(k,t),\nabla_k P^\epsi_N(k,t)] \) +\Or(\epsi^{N})\,,
\end{eqnarray*}
which proves formula \eqref{mres} in Theorem~\ref{th1}.

To conclude \eqref{poleps}, we observe that evaluating the traces in the definitions of
the curvatures $\Theta(k,t)$ and $\Theta_N^\epsi(k,t)$ with respect to the orthonormal basis $(\chi_\alpha(k,t))_{\alpha=1}^M$ of $\ran P(k,t)$
resp.\ the orthonormal basis $(\chi^\epsi_\alpha(k,t))_{\alpha=1}^M = (T^{\epsi *}_N(k,t)\, T_0(k,t)\chi_\alpha(k,t))_{\alpha=1}^M$ of $\ran P^\epsi_N(k,t)$ yields
\begin{align*}
    \Theta (k,t)  =  & \ - \I \tr \(P(k,t) \, [\d_t P(k,t), \, \nabla_k P(k,t)] \)\\
    =  & \ 2   \I \sum_\alpha  \im \< \d_t \chi_\alpha(k,t), \nabla_k \chi_\alpha(k,t) \>
\end{align*}
and likewise for $\Theta_N^\epsi(k,t)$,
\begin{align*}
    \Theta^\e_N (k,t)  =  & \ - \I \tr \(P^\e_N(k,t) \, [\d_t P^\e_N(k,t), \, \nabla_k P^\e_N(k,t)] \) \\
     =  & \ 2   \I \sum_\alpha  \im \< \d_t \chi^\e_\alpha(k,t), \nabla_k \chi^\e_\alpha(k,t) \>\,.
\end{align*}
Thus we  obtain $ \Theta(k,t) = - \partial_t \A (k,t) - \nabla_k \phi (k,t) $,
where the geometric vector potentials is
\begin{equation*}
\A(k,t)=  \I \sum_\alpha \< \chi_\alpha(k,t) , \nabla_k \chi_\alpha  (k,t) \> \,,
\end{equation*}
and the geometric scalar potential is
\begin{equation*}
\phi(k,t) = - \I \sum_\alpha \< \chi_\alpha(k,t) , \d_t \chi_\alpha (k,t) \>\,.
\end{equation*}
The decomposition for $\Theta^\epsi_N(k,t)$ in terms of $\A^\epsi_N(k,t)$ and $\phi^\epsi_N(k,t)$ is then completely analogous with
$\chi^\e_\alpha(k,t)$ replacing $\chi_\alpha(k,t)$.
By construction we have that $T^\epsi_N(k,0)=T_0(k,0)$ and $T^\epsi_N(k,T)=T_0(k,T)$ and therefore
$\chi_\alpha(k,0)=\chi_\alpha^\epsi(k,0)$, as well as
$\chi_\alpha(k,T)=\chi_\alpha^\epsi(k,T)$. As a consequence also $\A^\e_N(k,0)=\A(k,0)$ and $\A^\e_N(k,T)=\A(k,T)$.
Hence
\begin{eqnarray*}
 -\int_{0}^{T} \!\!\!\D t \int_{\T^*} \D k\, \Theta^\epsi_N (k,t)
&=&  \int_{0}^{T} \!\!\!\D t \int_{\T^*} \D k\,  (\partial_t \A^\epsi_N (k,t) + \nabla_k \phi^\epsi_N (k,t) )  \\
&= &  \int_{\T^*} \D k\, \left(  \A^\epsi_N (k,T) -  \A^\epsi_N (k,0) \right) \\
& = &\int_{\T^*} \D k\, \left(  \A  (k,T) -  \A  (k,0) \right)   \\
& =  & -\int_{0}^{T} \!\!\!\D t \int_{\T^*} \D k\, \Theta  (k,t) \,,
\end{eqnarray*}
where the second equality follows from the periodicity of $\phi^\epsi_N(k,t)$ in $k$ and for the last equality one just reverses the preceding steps for the adiabatic instead of the superadiabatic quantities.
$\Box$

\subsection{Proof of Theorem \ref{th2} }

In the proof of our second result we will make heavy use of the semiclassical calculus for operators
with equivariant symbols as presented in \cite{PST, Te}. The reader is referred to these references for
more details as we shall hereafter use the developed semiclassical techniques without any further ado.

In the following we consider the case where all Bloch bands within $\ran P(k,t)$, $t\in I$, are isolated and
non-degenerated. It suffices then to restrict ourselves to only one of these bands.
Thus, the results of Section~\ref{super} yield an effective Hamiltonian $H_{{\rm eff}}(k,t)$,
acting as a simple multiplication operator on the reference space $L^2(\T^*,\C)$, such that
\begin{equation} \label{symbol}
H_{{\rm eff}}(k,t)= E_m(k,t)+\O_{\mathcal E}(\e^2).
\end{equation}
The corresponding classical equations of motion are simply given by
\begin{equation}
\label{ham1}
\left \{
\begin{aligned}
& \dot r = \nabla_k E_m(k,t),\\
& \dot k = 0,
\end{aligned}
\right.
\end{equation}
and the generated classical flow will be denoted by
\begin{equation}\label{flow}
\widetilde \Phi_m(t,0):(k, \, r)\mapsto \left(k, \, r+\int_0^t \D s \, \nabla_k E_m(k,s)  \right).
\end{equation}
To conclude the proof of Theorem~\ref{th2} we need one more additional result, namely
the following Egorov-type theorem for time-dependent Hamiltonians.
\begin{lemma} \label{egorov}
Let $E_m(t)$ be an isolated and non-degenerated Bloch band for all $t \in I$.
Denote by
\begin{equation*}\label{ueff}
U^\e_{m}(k,t) = \exp\left(-\frac{\I}{\e} \int_0^t \D s \, E_m(k,s)\right)
\end{equation*}
the unitary propagator associated to $E_m(k,t)$.
Then, for any  $a\in C_{\rm b}^\infty$ there is a $C_a<\infty$ such that
\begin{equation}
\norm{\, U_{m}^\e(k,t)^*\,  a^W \, U_{m}^\e(k,t)- {(a \circ \widetilde\Phi_m(t,0))}^W }\leq \e^2 C_a |t|(1+|t|),
\end{equation}
where $\widetilde \Phi_m(t,0)$ is given by \eqref{flow}.
\end{lemma}
\begin{proof} The proof is almost analogous to the time-independent case, nevertheless it is given here for completeness.
First note that the function $E_m(k,s)$ is bounded together with
its partial derivatives (on bounded time intervals). Having in mind \eqref{flow}, we note
that $(a \circ \widetilde \Phi_m (t,0))\in C_{\rm b}^\infty $, as well as
$\partial_t (a \circ \widetilde \Phi_m (t,0))\in C_{\rm b}^\infty$, for any fixed $t\in \R$.
We therefore can interchange
quantization and differentiation w.r.t. $t\in \R$ and write
\begin{equation*}
\begin{aligned}
& \, U_{m}^\e(k,t)^* \,  a^W  \, U_{m}^\e(t,k)- {(a \circ \widetilde\Phi_m(t,0))^W} = \\
& \, = \int_{0}^t {{\rm d}s} \, \frac{{\rm d}}{{\rm d}s} \left(U_{m}^\e(k,s)^* \, {(a \circ \widetilde \Phi_m(t,s))^W}
 \, U_{m}^\e(s,k)\right) \\
& \, =  \int_{0}^t  {{\rm d}s} \, U_{{m}}^\e(k,s)^* \, I(t,s) \, U_{{m}}^\e(k,s) \, .
\end{aligned}
\end{equation*}
where we denote
$$
I(t,s)\equiv \frac{\I}{\e} \, \big[E_m(k,s),  {(a \circ \widetilde \Phi_m(t,s))^W}\big] +
\frac{{\rm d}}{{\rm d}s} \, (a \circ \widetilde \Phi_m(t,s))^W .
$$
Having in mind \eqref{flow}, we easily get
$$
\frac{{\rm d}}{{\rm d}s} \, {(a \circ \widetilde \Phi_m(t,s))}=
-\nabla_{k} E_m(k,s) \cdot \nabla_{r}  (a \circ \widetilde \Phi_m(t,s)).
$$
Thus we can use Moyal's expansion of the commutator, to obtain
$$
\frac{\I}{\e} \, \big[E_m(k,s),  (a \circ \widetilde \Phi_m(t,s)) \big]_{\sharp}  -
 \nabla_{k} E_m(k,s) \cdot \nabla_{r}  (a \circ \widetilde \Phi_m(t,s))= \O(\e^2),
$$
where $\sharp$ denotes the Moyal product of symbols. Moreover, by using the
the simple time-dependence of the classical flow \eqref{flow}, we also obtain that
the integrand $I(t,s)$ is indeed $\O(\e^2(1+|t-s|))$, since derivatives w.r.t $k$
of $(a \circ \widetilde \Phi_m(t,s))$ grow linearly in time for large enough $t$. This fact together with the consequent
integration in time of $I(t,s)$ then proves assertion of the lemma.
\end{proof}
\begin{proof}[Proof of Theorem~\ref{th2}] The above given lemma is stated in terms of observables on the reference space $L^2(\T^*,\C)$.
In order to obtain the analogous result in the physical phase space and the correpsonding
effective semiclassical equations of motion \eqref{sceq} we have to undo both,
the mapping to the reference space $T_N^\e$ as well as the Bloch-Floquet transformation $\F$.
To this end, Moyal's expansion will allow us to derive explicit formulas up to sufficient high orders in $\e$.

Let us first study how the Bloch-Floquet transformation $\F$
maps observables on $\mathcal H= L^2(\R^d)$ to observables in the corresponding Zak representation.
From \cite{PST}, we know that
\be\label{zmapping}
{\F}^{-1} \, a^W( \I \epsi \nabla_k^\tau, k) \, {\F} = a^W(\epsi x, -\I \nabla_x)
\ee
whenever $a(r,k)\equiv a(r,k+\gamma^*)$ $\forall$ $\gamma \in \Gamma^*$.
Here $a^W(\epsi x, -\I \nabla_x)$ denotes the Weyl quantized operator
acting on $L^2(\R^d,\C)$, whereas the operator $a^W( \I \epsi \nabla_k^\tau, k)$ acts on $L^2(Y^*, L^2(\T^d))$.
One should note confuse these two types of quantization,
even though they are both obtained from the same symbol $a(r,k)$. Moreover $a^W( \I \epsi \nabla_k^\tau, k)$, \ie
observables in the Zack representation, should be distinguished from observables
$a^W(\I \e \nabla_k, k)$ acting on the reference space $L^2(\T^*,\C)$.

Next we define a change of coordinates
(for each fixed $t\in \R$) by
\begin{equation}
\Sigma^\e_t: \R^{2d}\rightarrow \R^{2d},\quad (r,k)\mapsto (r+\e \mathcal A_m(k,t), k),
\end{equation}
where   $\mathcal A_m(k,t)$ is the  Berry connection, as defined in (\ref{Berry connection}).
We claim   that the unitary operator $T_N^\e(t):P^\epsi_N(t)\,L^2(Y^*, L^2(\T^d)) \to L^2(Y^*)$
constructed in Proposition~\ref{adia} maps semiclassical
observables in the Bloch-Floquet representation to observables
in the reference space via
\begin{equation}
\label{mapping}
T_N^\e( t) \, a^W \,   {T_N^\e}^{*}(t) =  {((a \circ \Sigma^\e_t) (k,r))}^W + \O(\e^2),
\end{equation}
where here and in the following  $\O(\e^n)$ refers to the norm of bounded operators.
This formula should be compared to \eqref{zmapping}: Whereas $\F$ leaves the semiclassical symbol $a(r,k)$ invariant, the
unitary mapping to the reference space $T_N^\e$ does not. This can be seen by using   Moyal's expansion, \ie we have to expand
$$
T_N^\e (k,t) \, \sharp \, a(r,k) \, \sharp \, {T_N^\e}^{*}(k,t)
$$
in powers of $\e$. To this end recall that we
explicitly constructed $T_N^\e = T_0 + \e \, T_1 + \O_{\mathcal E}(\e^2)$
in the proof of Proposition~\ref{adia} above and that the equivariance property of $T^\e_N(k,t)$ assures that we can interpret $T^\e_N(k,t)$  as an operator valued semiclassical symbol in the sense of \cite{PST,Te} with quantization $T^\e_N(t)$.
Since $a(r,k)$ is scalar-valued, a straightforward calculation yields
\begin{align*}
T_N^\e (k,t) \, \sharp \, a(r,k) \, \sharp \,  {T_N^\e}^{*}(k,t) = & \ a(r,k) +   \I \e \,
\nabla_r a(r,k) \cdot T_0(k,t)\nabla _k T^*_0(k,t) \\
& \ + \O (\e^2),
\end{align*}
where we have used that  $T_N^\e(k,t)$ does \emph{not} depend on $r$ and the fact that
$$
T_0(k,t) T_1^*(k,t) + T_1(k,t) T_0^*(k,t)=0,
$$
by relation \eqref{rela}. A comparison with the Taylor expansion in powers of $\e$ for
$a \circ \Sigma^\e_t$ proves the claim.

Combining \eqref{zmapping} and \eqref{mapping} allows us to transform the Egorov theorem \ref{egorov} into the
corresponding result on the original Hilbert space $L^2(\R^d)$ and thus to conclude the assertion of Theorem~\ref{th2}.
To this end we write
\begin{align*}
P(0) \, U^\e(t)^*\,  a^W \, U^\e(t) \, P(0) =
\F^{-1} P_N^\e(0) \, U^\e(t)^*\,  a^W \, U^\e(t) \, P_N^\e(0) \F .
\end{align*}
Invoking Proposition~\ref{adia} we obtain
\begin{align*}
& P_N^\e(0) \, U^\e(t)^*\,  a^W \, U^\e(t) \, P_N^\e(0)=\\
& \ =  \, T_N^\e(0) \, U_{{\rm a}}^\e(t)^* \,  a^W \, U_{{\rm a}}^\e(t) \, T_N^{\e *}(0) + \O(\e^N) \\
& \ =   \, U_{{m}}^\e(t)^* \, T_N^\e (t)^* \,  a^W \, T_N^\e (t)^* \, U_{{m}}^\e(t) \, + \O(\e^N)\\
& \ =   \, U_{{m}}^\e(t)^* \, {(a \circ \Sigma^\e_t)}^W \, U_{{m}}^\e(t) +   \O(\e^2),
\end{align*}
where for the last equality we simply inserted \eqref{mapping}. By Lemma~\ref{egorov}, this yields
\begin{align*}
& \, P_N^\e(0) \, U_m^\e(t)^*\, {(a \circ \Sigma^\e_t)}^W \, U_m^\e(t) \, P_N^\e(0)  = \\
& \, =   \, P_N^\e(0) \, {((a \circ \Sigma^\e_t) \circ
 \widetilde \Phi_m^\e(t,0))}^W \, P_N^\e(0) +   \O(\e^2|t|(1+|t|))\\
& \, =   \, P_N^\e(0) \, {\left((b  \circ \Phi_m^\e(t,0)\right)\circ \Sigma^\e_t) }^W \, P_N^\e(0) +   \O(\e^2|t|(1+|t|)),
\end{align*}
where we define the classical flow, which maps observables on physical phase space, via
$$
\Phi_m^\e(t,0):=\Sigma_t\circ \widetilde \Phi_m^\e(t,0) \circ \Sigma_t^{-1}.
$$
Thus we denote the new coordinates $(q,p)\in \R^{2d}$ by
$$
q=r+\e \mathcal A_m(k,t), \quad p = k.
$$
Using \eqref{ham1}, it is then straightforward to
express Hamilton's equations of motion (in the $m$th Bloch band) in these new coordinates, \ie
\begin{equation*}
\dot q = \nabla_p E_m(p,t) -   2 \e \im \< \partial_t \ph_m(p,t), \nabla_p \ph_m(p,t)\>,
\qquad \dot p = 0,
\end{equation*}
and since, as before,
\begin{equation*}
\im  \< \d_t \ph_m (k,t), \nabla_k \ph_m (k,t) \>
=  - \frac{\I}{2} \tr \(P_m(k,t) \, [\d_t P_m(k,t), \, \nabla_k P_m(k,t)] \),
\end{equation*}
we clearly get \eqref{sceq}. In summary this yields
\begin{align*}
\norm{ \, P_m(0)  \left(U^\e(t)^*\,  a^W \, U^\e(t)  -
  {\left(a  \circ \Phi_m^\e(t,0)\right) }^W \right)  P_m(0)} \leq   C  \e^2|t|(1+|t|),
\end{align*}
or, in other words, for any $\psi_0 \in \ran P_m(0)$ we get
\begin{equation*}
\left | \, {\langle\psi_0, U^\e(t)^*\,  a^W \, U^\e(t) \, \psi_0\rangle}_{L^2} -
{\langle \psi_0, {\left(a  \circ \Phi_m^\e(t,0)\right) }^W \, \psi_0\rangle}_{L^2} \right|\leq C \e^2 |t|(1+|t|).
\end{equation*}
Finally, by using identity \eqref{wigner2}, we convert this Egorov type theorem for operators into
the analogous one for Wigner functions, \cf \cite{TePa}, having in mind that $\Phi^\e_m$ is volume preserving.
Thus Theorem~\ref{th2} is proved.
\end{proof}

\section{Symmetries and geometric interpretation of currents}\label{sgeo}

\subsection{Symmetries}
In many physical problems, the role of symmetries is crucial for a deep
understanding of the dynamics. Piezoelectricity is no exception. Indeed we prove that
 the piezoelectric current is zero \emph{if
space-reflection symmetry is not broken}, in agreement with the common lore in solid
state physics.

As usual, space-reflection symmetry is realized in $\mathcal H=L^2(\R^d)$ by the operator $\mathscr R$, defined by
\[
\( \mathscr R \psi \)(x) = \psi(-x), \qquad  \psi \in L^2(\R^d).
\]
The group structure of the periodicity lattice $\Gamma$ implies that $- \Gamma =
\Gamma$, \ie that for any $\gamma \in \Gamma$ one has $- \gamma \in \Gamma$.
Therefore $[-x] = - [x] \in Y$, where we introduce $x=[x]+\gamma$ for the a.e.
unique decomposition for $x\in \R^d$ as a sum of $[x]\in Y$ and $\gamma\in \Gamma$.
Equipped with this observation it is easy to check that $\widetilde{\mathscr R}= \F
\mathscr R \F^{-1}$ acts as
\[
\( \widetilde{\mathscr R} \psi \)(k,y) = \psi(-k,-y), \qquad  \psi \in L^2(Y_k^{*}) \otimes \Hf,
\]
or equivalently $(\widetilde{\mathscr R} \psi)(k) =
\mathscr R_{\rm f} \psi(-k)$ where $\mathscr R_{\rm f}$ is the space reflection operator in $\Hf = L^2(\T_y^d)$.

\goodbreak

Notice that, if $H(t) = - \frac{1}{2}\, \Delta + V_{\Gamma}(t,x)$, then
the condition $[H(t), \mathscr R]= 0$ is fulfilled whenever $V_{\Gamma}(t, -x) = V_{\Gamma}(t, x)$. Some authors refer to this condition by
saying that \emph{the crystal has a center of inversion}. However, the use of the world ``crystal'' should not obscure the fact that
the latter is a property of $V_{\Gamma}$, not a property of $\Gamma$.

\begin{proposition}[\textbf{Space-reflection symmetry}] \label{Prop spacerefl} Assume that the self-adjoint operator $H(t)$ commutes with $\mathscr R$,
and that $\F H(t) \F^{-1}$ is a \emph{continuously} fibered operator.
Let $P(\cdot, t)$ be either
\renewcommand{\labelenumi}{{\rm(\alph{enumi})}}
\begin{enumerate}
    \item $P(\cdot, t) = \1_{(-\infty, E(t))}(H(\cdot,t))$, or  \vspace{2mm}
    \item $P(\cdot, t) = P_m(\cdot, t)$ the eigenprojector corresponding to an
    isolated Bloch band $E_m(\cdot,t)$.
\end{enumerate}
Then
\begin{equation}\label{Spacerefl projector}
P(k,t) =  \mathscr R_{\rm f} \, P(-k,t) \, \mathscr R_{\rm f},
\end{equation} and the piezocurvature $\Theta = - \I \tr \(P \, [\d_t P, \, \nabla_k
P] \) $ satisfies \begin{equation} \label{Spacerefl curvature} \Theta(-k,t) = - \,
\Theta(k,t).
\end{equation}
\end{proposition}
\begin{proof} The transformed Hamiltonian $\F H(t) \F^{-1}$ commutes with $\widetilde{\mathscr R}$, yielding a symmetry of the fibers,
\ie
\be \label{Spacerefl hamiltonian}
H(k,t) = \mathscr R_{\rm f} \, H(-k,t) \, \mathscr R_{\rm f}.
\ee
In order to prove (\ref{Spacerefl projector}) one distinguish two cases.

Case (a). Since by assumption $E(t) \in \varrho \(H(t)\)$, the resolvent set of $H(t)$, one has $E(t) \in \varrho\(H(k,t)\)$
for a.e. $k \in Y^*$. By the continuity of the fibration the same holds true for every
$k \in Y^*$. Then, by applying functional calculus to (\ref{Spacerefl hamiltonian}), one obtains
(\ref{Spacerefl projector}).

Case (b). By the unitary equivalence (\ref{Spacerefl hamiltonian}) $E_m(k,t)$ is an eigenvalue
of $H(-k,t)$.  By the \emph{continuity} of $k \mapsto E_m(k,t)$ and the \emph{gap condition}, by starting from $k=0$
one concludes that $E_m(-k,t) = E_m(k,t)$ for any $k$, \ie $E_m(k,t)$ is the eigenvalue corresponding to
$P(-k,t)$. Let $f \in C^{\infty}_0(\R)$ be the smoothed characteristic function of an interval containing $E_m(k,t) =
E_m(-k,t)$ and no other point of  $\sigma\(H(k,t)\) = \sigma\(H(-k,t)\)$. Then from (\ref{Spacerefl hamiltonian}) one gets
\[
  P(k,t) = f(H(k,t)) = \mathscr R_{\rm f} f(H(-k,t)) \mathscr R_{\rm f} = \mathscr R_{\rm f} P(-k,t) \mathscr R_{\rm f}.
\]

In both cases from (\ref{Spacerefl projector}) one computes
\begin{align*}
\d_{k_i} P(k,t)  =  - \, \mathscr R_{\rm f} \, \d_{k_i} P(-k,t) \, \mathscr R_{\rm f}, \qquad
\d_t P(k,t)  =   \mathscr R_{\rm f} \, \d_t P(-k,t)  \, \mathscr R_{\rm f},
\end{align*}
so that
\begin{align*}
\I \, \Theta(-k,t)  = &  \ - \tr \( \mathscr R_{\rm f} P(k,t) \mathscr R_{\rm f} \, \mathscr R_{\rm f} [\d_t P(k,t), \, \nabla_k P(k,t)] \mathscr R_{\rm f}  \) \\
              = &  \ - \tr \( P(k,t) [\d_t P(k,t), \, \nabla_k P(k,t)] \) \\
              = &  \ - \I \, \Theta(k,t).
\end{align*}
\end{proof}

\goodbreak


We now turn to study the consequences of time-reversal symmetry. This symmetry
is realized in $\mathcal H=L^2(\R^d)$ by the complex conjugation operator, \ie by the operator
\[
( \mathscr C \psi )(x) = \bar{\psi}(x), \qquad  \psi \in L^2(\R^d).
\]
By the Bloch Floquet transform we get that $\widetilde{\mathscr C}= \F \mathscr C \F^{-1}$ acts as
\[
( \widetilde{\mathscr C} \psi )(k) = \mathscr C_{\rm f} \, \psi(-k),  \qquad \psi \in L^2(Y_k^{*}) \otimes \Hf,
\]
where $\mathscr C_{\rm f}$ is the complex conjugation operator in $\Hf$.
\begin{proposition}[\textbf{Time-reversal symmetry}] \label{Prop timerev}
Assume that the self-adjoint operator $H(t)$ commutes with $\mathscr C$ in
$L^2(\R^d)$, and that $\F H(t) \F^{-1}$ is a \emph{continuously} fibered operator.
Let $P(\cdot, t)$ be as in Proposition \ref{Prop spacerefl}.
Then
\be \label{Timerev projector}
P(k,t) = \mathscr C_{\rm f} \, P(-k,t) \, \mathscr C_{\rm f},
\ee
and the piezocurvature $\Theta = - \I \tr \(P \, [\d_t P, \, \nabla_k P] \) $ satisfies
\be \label{Timerev curvature}
\Theta(-k,t) =  \Theta(k,t).
\ee
\end{proposition}

The difference in sign between (\ref{Timerev curvature}) and (\ref{Spacerefl curvature}) is due to the fact that
$\mathscr C_{\rm f}$ is an antilinear unitary operator, while $\mathscr R_{\rm f}$ is a linear one.
Notice that the time dependence plays no role in the statement and in the proof of this proposition.
Since the proof is very similar to the one of Proposition \ref{Prop spacerefl} we only point out
some differences due to the fact that $\mathscr C_{\rm f}$ is an\emph{ antilinear} operator.

\begin{proof} The starting point is again a symmetry of the fibers, namely
\begin{equation}
\label{Timerev hamiltonian}
H(k,t) = \mathscr C_{\rm f} H(-k,t)\mathscr C_{\rm f}.
\end{equation}

Case (a). One proceeds as in the previous proof, exploiting the fact that functional calculus is
covariant with respect to complex conjugation, \ie $f(\mathscr C_{\rm f} \, A \, \mathscr C_{\rm f}) =
\mathscr C_{\rm f} \, f(A) \,\mathscr C_{\rm f}$ whenever $A$ is self-adjoint
and $f$ is an admissible function.

Case (b). By assumption there exists $\ph_m \in \Hf, \, \ph_m \neq 0$, such that
\[
H(k,t) \ph_m = E_m(k,t) \ph_m
\]
By complex conjugation one gets
\[
E_m(k,t) \, \mathscr C_{\rm f}\ph_m = \mathscr C_{\rm f} H(k,t) \ph_m = \mathscr C_{\rm f} H(k,t) \mathscr C_{\rm f} \,
\mathscr C_{\rm f} \ph_m = H(-k,t) \, \mathscr C_{\rm f} \ph_m,
\] which shows that $E_m(k,t)$ is an eigenvalue of $H(-k,t)$. Then one concludes the argument as in the
previous Proposition.

In both cases by (\ref{Timerev projector}) one has
\begin{equation*}
\d_{k_i} P(k,t)  =   - \, \mathscr C_{\rm f} \, \d_{k_i} P(-k,t) \, \mathscr C_{\rm f}, \quad
\d_t P(k,t)  =   \mathscr C_{\rm f} \, \d_t P(-k,t)  \, \mathscr C_{\rm f},
\end{equation*}
so that
\begin{align*}
\I \, \Theta(-k,t)  = & \  - \tr \( \mathscr C_{\rm f} P(k,t) \mathscr C_{\rm f} \,
\mathscr C_{\rm f} [\d_t P(k,t), \, \nabla_k P(k,t)] \mathscr C_{\rm f}  \) \\
                = &  \ \tr \( P(k,t) [\d_t P(k,t), \, \nabla_k P(k,t)] \) \\
                = &  \ \I \, \Theta(k,t),
\end{align*}
where the minus sign disappears because of the \emph{antilinearity} of $\mathscr C_{\rm f}$.
\end{proof}

\subsection{Geometric reinterpretation and electromagnetic analogy}
It is worthwhile to comment the relationship between the piezocurvature
$\Theta$ and the \emph{curvature of the Berry connection $\Omega$}. The latter is the differential 2-form
over $\T^*$ whose components are given by
\[
\Omega_{j,l}(k,t) =  -\I  \tr \( P(k,t)  \, [\d_j P(k,t), \, \d_l P(k,t)] \), \qquad  j,l \in\{1,\dots, d \}
\]
where $\d_j = \d / \d k_j$. As usual, this 2-form is identified with a vector field when convenient.\footnote{
\quad In dimension $d =3$ one identifies the antisymmetric matrix $\Omega_{i,j}$ with the
vector with components $\Omega_l = \sum_{i,j} \epsi_{lij} \Omega_{i,j}$, where $\epsi_{ijl}$ is the totally antisymmetric symbol.
Then $\Omega$ can be considered a vector field over $\T^3$. Similarly, for $d=2$ the 2-form $\Omega$ is identified with a
scalar function over $\T^2$.} 
The importance of $\Omega$ is well-known. This curvature is indeed the main tool to compute the Chern class of the Bloch bundle
\ie the complex vector bundle over $\T^*$ whose fiber at the point $k \in \T^*$ is the subspace generate by
a relevant set of Bloch functions. The vanishing of the mentioned Chern class is crucial in
order to prove existence of localized Wannier functions
\cite{Ne83, MaVa, Pa}
while in the magnetic case the analogous quantity represents the quantized transverse
conductivity in the Integer Quantum Hall Effect \cite{Bellissard}.

In the following $P(k,t)$ can be interpreted in both the senses mentioned in Proposition \ref{Prop spacerefl}.
It is convenient to introduce Greek indexes $\mu, \nu \in \{0, 1,\ldots, d \}$ and to pose $k_0 =t$ and
$\d_0 = \d/\d t$. The differential 2-form
\begin{align*}
  \Xi(k,t)  :=& \, \sum_{\mu, \nu =0}^{d}  \Xi_{\mu, \nu}(k,t) \,  dk_{\mu} \wedge dk_{\nu} \\
  \Xi_{\mu,\nu}(k,t) =& \,   -\I  \tr \( P(k,t)  \, [\d_\mu P(k,t), \, \d_\nu P(k,t)] \)
\end{align*}
over $M := \T^d \times \R$ represents the curvature of a connection on the \emph{extended Bloch bundle},
\ie the complex vector bundle over $M$ whose fiber at $(k,t)$ is the range of $P(k,t)$.
More specifically $\Xi$ is the curvature of the connection induced by the trivial connection in the trivial bundle (\ref{trivial}),
which is sometimes called \emph{induced connection}.
Equipped with this notation one has
\[
\Omega_{\, j,l}(k,t) = \Xi_{ j,l}\,(k,t) \quad \mbox{and}  \quad \Theta_{j}(k,t) = \Xi_{ j,0}\, (k,t),
\]
where $j,l \indexd$.

The analogy with the electromagnetic field is striking: in this analogy $\Theta$ is identified with the electric field,
while $\Omega$ with the magnetic field. The two fields combine into a tensor over "space-time" $M$. This analogy,
  rooted in the fact that both electromagnetism and the gauge theory of Bloch bands are gauge theories with structure group $U(1)$,
is reinforced by an analysis of the transformation properties under time-reversal and space-reflection symmetry, which are summarized in the following self-explanatory table.

\[
\begin{array}{ccccc}
\mbox{\textbf{Quantity}}           & \quad \quad & \mbox{\textbf{Time-reversal}}    & \quad \quad &  \mbox{\textbf{Space-reflection}}  \\
                                   & \quad \quad & \mbox{\textbf{symmetry}}    & \quad \quad &  \mbox{\textbf{symmetry}}  \\[5 mm]
   P(-k) = & \quad &   \mathscr C_{\rm f} \, P(k) \, \mathscr C_{\rm f}   & \quad &       \mathscr R_{\rm f} \, P(k) \, \mathscr R_{\rm f}           \\[3 mm]
   \A(-k)=                & \quad &     + \A(k)                      & \quad&           - \A(k)                  \\[3 mm]
 \Omega(-k) =             & \quad &     - \Omega(k)                  & \quad&         + \Omega(k)                \\[3 mm]
 \Theta(-k) =             & \quad &     + \Theta(k)                  & \quad&       - \Theta(k)                \\[3 mm]
\end{array}
\]

The table shows clearly that breaking of space-reflection symmetry is a necessary condition in order to
have a non-zero piezoelectric current, while breaking of time-reversal symmetry is necessary in order to
have a non-zero Chern class of  the Bloch bundle over $\T^*$ and consequently a non-zero Hall conductance \cite{Bellissard, TePa}.

We now specialize to case (b), namely $P(k,t)= P_m(k,t)$ is the eigenprojector corresponding to
an isolated Bloch band.  Then by introducing the Berry connection (\ref{Berry connection}) and the geometric scalar potential (\ref{geometric potential}),
one obtains (locally in $k$) the following expressions,
\[
\begin{array}{lcl}
  \Omega_m = \nabla \wedge \A_m                       & \quad \quad &                \Omega_{m \, j,l} = \d_j \A_{m \, l} - \d_l \A_{m \, j}      \\[3 mm]
  \Theta_m = -\d_t \A_m - \nabla \phi_m               & \quad \quad &                \Theta_{m \, j}   =  -\d_t \A_{m \, j} - \d_j \phi_m.         \\
\end{array}
\]
The latter formulae show again a strict analogy with the electromagnetic field, as  noticed in \cite{KoTa}.

\begin{remark}\textbf{(The case of a periodic deformation)}
The analysis of the previous sections, including formula (\ref{mres}), extends to
the case of a periodic deformation of the crystal, \ie
\[
V_{\Gamma}(\cdot, t + T) = V_{\Gamma} (\cdot, t) \quad  \forall \, t \in \R,
\]
for a suitable period $T$. In the periodic case the extended Bloch bundle $\xi$ can
be regarded as a vector bundle over $ B_{\rm per}:= \T^* \times \T_t^1$. Since
$B_{\rm per}$ is a boundaryless manifold, the theory of characteristic classes
applies to this case. In particular, focusing on $d=3$ and choosing the third
direction for sake of definiteness, the quantity
\[
C_3(k_1,k_2) :=  \, \frac{1}{2\pi}  \int_0^T \D t \int_{\T^1} \D k_3 \,
\Theta(k,t)_3,
\]
where $k = (k_1,k_2,k_3)$ and $\T^1$ is the circle in the direction $\gamma_3^* \in
\Gamma^{*}$, corresponds to the first Chern class of the restriction of $\xi$ to the
sub-manifold $\T^1_{k_3} \times \T^1_t \subset B_{\rm per}$, and as such is an
integer. Since this integer depends continuously on $(k_1,k_2)$, it is constant, \ie
$C_3(k_1,k_2) =c_3 \in \Z$.

On the other side, the same argument shows that the quantity
\[
\tilde{C}_3(k_1,k_2) := \frac{1}{2\pi} \, \int_0^T \D t \int_{\T^1} \D k_3 \,
\Theta_N^{\epsi}(k,t)_3,
\]
is also an integer, denoted as $\tilde{c}_3$. Thus from $\Theta_N^{\epsi}(k,t) =
\Theta(k,t) + \O(\e)$ and $\e \ll 1$ one concludes that $\tilde{c_3} = c_3$, in
agreement with the so-called \emph{topological robustness} of the Chern class. In
view of that, one gets
\begin{eqnarray*}
  \Delta \! \p^\epsi &=&  -\frac{1}{(2\pi)^3} \, \int_{0}^{T} \!\!\!\D t \int_{\T^*} \D k\, \, \Theta^{\e}_N(k,t) +\Or(\epsi^{N})\\
                  &=&  -\frac{1}{(2\pi)^3} \, \int_{0}^{T} \!\!\!\D t \int_{\T^*} \D k\, \, \Theta(k,t) +\Or(\epsi^{N})
\end{eqnarray*}

which shows that the formula (\ref{poleps}) holds true in the periodic case, even if
$\dot H$ does not vanish at the boundaries of the time interval $[0,T]$. One
concludes that
\[
{(\Delta \! \p^\epsi)}_3 =   - \frac{1}{(2 \pi)^2}  c_3 + \Or(\e^N), \qquad c_3 \in
\Z.
\]
The macroscopic polarization is quantized up to errors smaller than any power of
$\e$. As for $d=1$, this fact has been pointed out by Thouless \cite{Th}. Notice
that, in the adiabatic limit, the relative polarization is essentially a geometric
quantity, namely the Chern class of a vector bundle. In particular it does not
depend on the speed of the periodic deformation, \ie $H(t)$ and $H(\lambda t)$,
$\lambda \in \R$, lead to the same macroscopic polarization.
\end{remark}

\bibliographystyle{amsplain}

\begin{thebibliography}{9999}

\bibitem{ABL} J.~E.\ Avron, J.\ Berger, and Y.\ Last, \emph{Piezoelectricity: quantized charge transport
driven by adiabatic deformations}, Phys.\ Rev.\ Lett.\ {\bf 78} (1997), 511--514.

\bibitem{Bellissard} J.~Bellissard, A.~van Elst, H.~Schulz-Baldes, \emph{The noncommutative geometry of the quantum Hall effect},
 J.  Math. Phys. {\bf 35} Issue 10, 5373--5451 (1994).


\bibitem{ElSch} A.\ Elgart and B.\ Schlein, \emph{Adiabatic charge transport and the Kubo formula
for Landau type hamiltonians},  Comm.\ Pure and Appl.\ Math.\ {\bf 57} (2004), 590--615.

\bibitem{FaZi} F. Faure, B. Zhilinskii, \emph{Topological properties of the
Born-Oppenheimer approximation and implications for the exact spectrum},  Letters in
Math. Physics {\bf 55}, 219-238, (2001).

\bibitem{Fol} G. Folland, \emph{Harmonic Analysis in Phase Space}, Princeton University Press (1989).

\bibitem{Ka} T. Kato, \emph{On the adiabatic theorem of quantum mechanics}, Phys. Soc. J. {\bf 5} (1950), 435--439.

\bibitem{KSV} R.~D. King-Smith and D. Vanderbilt, \emph{Theory of polarization of crystalline solids},
Phys. Rev. B {\bf 47} (1993), 1651--1654 .

\bibitem{KlSe} M.\ Klein and R.\ Seiler, \emph{Power-law corrections to the Kubo formula vanish in quantum Hall systems},
Commun.\ Math.\ Phys.\ {\bf 128} (1990), 141--160.

\bibitem{KoTa} H. Koizumi and Y. Takada, \emph{Geometric phase current in solids: Derivation in a path integral
approach}, Phys. Rev. B \textbf{65} (2002), 153104--153110.

\bibitem{Ma} R. M. Martin, \emph{Comment on calculations of electric polarization in crystals}, Phys. Rev.
B {\bf 9} (1974), 1998--1203.

\bibitem{MaVa} N. Marzari and D. Vanderbilt,  \emph{Maximally localized generalized Wannier functions for composite energy bands},
Phys. Rev. B {\bf 56}, 12847-12865 (1997)

\bibitem{Ne83} G.\ Nenciu, \emph{Existence of the exponentially localised Wannier
functions}, Comm. Math. Phys. {\bf 91}, Number 1, 81--85 (1983).

\bibitem{Ne} G. Nenciu, \emph{Linear adiabatic theory. Exponential estimates},
Comm. Math. Phys. \textbf{152} (1993), 479--496.

\bibitem{NTW} Q. Niu, D.~J. Thouless, and Y.~S. Wu, \emph{Quantized Hall conductance as a topological invariant},
Phys. Rev. B (3) {\bf 31} (1985), no. 6, 3372--3377.

\bibitem{PST1} G. Panati, H. Spohn, and S. Teufel, \emph{Space-Adiabatic Perturbation Theory},
Adv. Theor. Math. Phys. {\bf 7} (2003), 145--204.

\bibitem{PST} G. Panati, H. Spohn, and S. Teufel, \emph{Effective dynamics for Bloch electrons:
Peierls substitution and beyond}, Comm. Math. Phys. \textbf{242} (2003), 547--578.

\bibitem{Pa} G. Panati, \emph{Triviality of Bloch and Bloch-Dirac bundles}, preprint ArXive
{\tt math-ph/0601034}.

\bibitem{Re92} R. Resta, \emph{Theory of the electric polarization in crystals}, Ferroelectrics \textbf{136} (1992), 51--75.

\bibitem{Re94} R. Resta, \emph{Macroscopic polarization in crystalline dielectrics: the geometric phase approach}, Rev. Mod.
Phys. {\bf 66} (1994), no. 3, 899--915.

\bibitem{Resta_book1} R. Resta, \emph{Berry phase in electronic wavefunctions}, Lecture notes for the "Troisi\`{e}me
Cycle de la Phyique en Suisse Romande 1995-96" (1996).

\bibitem{Resta_book2} R. Resta, \emph{Berry phase and geometric quantum distance: macroscopic polarization and electron
localization}, Lecture notes for the "Troisi\`{e}me Cycle de la Phyique en Suisse Romande 1999--2000"
(2000).

\bibitem{Simon} B. Simon, \emph{Trace ideals and their applications}, London Math. Soc. Lecture Notes 35,
Cambridge University press, 1979.

\bibitem{Te} S. Teufel, \emph{Adiabatic perturbation theory in quantum dynamics},
Lecture Notes in Mathematics 1821, Springer (2003).

\bibitem{TePa} S. Teufel and G. Panati, \emph{Propagation of Wigner functions for the Schr\"{o}dinger equation with a perturbed periodic
 potential}, in "Multiscale Methods in Quantum Mechanics", eds. P. Blanchard and G. Dell'Antonio, Birkh\"{a}user 2004.

\bibitem{TKNN} D.\ J.\ Thouless, M.\ Kohomoto, M.\ P.\ Nightingale and M.\ den Nijs.
{\em Quantized Hall conductance in a two-dimensional periodic potential}, Phys.\ Rev.\ Lett.\ {\bf 49}, 405--408 (1982).

\bibitem{Th} D.~J. Thouless, \emph{Quantization of particle transport}, Phys. Rev. B {\bf 27} (1983), no. 10, 6083--6087.


\bibitem{XSN} D.~Xiao, J.~Shi, and Q.~Niu, \emph{Berry phase correction to electron density of states in solids},
Phys. Rev. Lett. {\bf 95} (2005), 137204--137208.

\bibitem {Wi} C.~H. Wilcox, \emph{Theory of Bloch waves}, J. d'Anal. Math. {\bf 33} (1978), 146--167.

\bibitem{Zak} J. Zak, \emph{Dynamics of Electrons in Solids in External Fields}, Phys. Rev. {\bf 168} (1968), 686-695.

\end{thebibliography}

\end{document}